\documentclass[aps,prd,amsmath,floats,floatfix,superscriptaddress,nofootinbib,showpacs]{revtex4}

\usepackage{amssymb}
\usepackage{amsmath}
\usepackage{verbatim}
\usepackage{mathrsfs}
\usepackage{amsfonts}
\usepackage{latexsym}
\usepackage{epsfig}
\usepackage{color}
\usepackage{graphicx,subfigure}
\usepackage{units}
\usepackage{overpic}
\usepackage{soul} 

\newcommand{\Real}{\mathbb{R}}

\newcommand{\re}{\mbox{Re}}
\newcommand{\im}{\mbox{Im}}

\begin{document}


\definecolor{orange}{rgb}{0.9,0.45,0} 

\newcommand{\argelia}[1]{\textcolor{red}{{\bf Argelia: #1}}}
\newcommand{\dario}[1]{\textcolor{red}{{\bf Dario: #1}}}
\newcommand{\juanc}[1]{\textcolor{green}{{\bf JC: #1}}}
\newcommand{\ayj}[1]{\textcolor{orange}{{\bf AyJ: #1}}}
\newcommand{\alberto}[1]{\textcolor{blue}{{\bf Alberto: #1}}}
\newcommand{\miguela}[1]{\textcolor{red}{{\bf MiguelA: #1}}}
\newcommand{\mm}[1]{\textcolor{orange}{{\bf MM: #1}}}
\newcommand{\OS}[1]{\textcolor{blue}{{\bf Olivier: #1}}}

\long\def\symbolfootnote[#1]#2{\begingroup%
\def\thefootnote{\fnsymbol{footnote}}\footnote[#1]{#2}\endgroup}


\title{On the linear stability of $\ell$-boson stars with respect to radial perturbations}

\author{Miguel Alcubierre}
\affiliation{Instituto de Ciencias Nucleares, Universidad Nacional Aut\'onoma de M\'exico,
Circuito Exterior C.U., A.P. 70-543, M\'exico D.F. 04510, M\'exico}

\author{Juan Barranco}
\affiliation{Departamento de F\'isica, Divisi\'on de Ciencias e Ingenier\'ias,
Campus Le\'on, Universidad de Guanajuato, Le\'on 37150, M\'exico}

\author{Argelia Bernal}
\affiliation{Departamento de F\'isica, Divisi\'on de Ciencias e Ingenier\'ias,
Campus Le\'on, Universidad de Guanajuato, Le\'on 37150, M\'exico}

\author{Juan Carlos Degollado}
\affiliation{Instituto de Ciencias F\'isicas, Universidad Nacional Aut\'onoma de M\'exico,
Apdo. Postal 48-3, 62251, Cuernavaca, Morelos, M\'exico}

\author{Alberto Diez-Tejedor}
\affiliation{Departamento de F\'isica, Divisi\'on de Ciencias e Ingenier\'ias,
Campus Le\'on, Universidad de Guanajuato, Le\'on 37150, M\'exico}

\author{Miguel Megevand}
\affiliation{Instituto de F\'isica Enrique Gaviola, CONICET. Ciudad Universitaria, 5000 C\'ordoba, Argentina}

\author{Dar\'io N\'u\~nez}
\affiliation{Instituto de Ciencias Nucleares, Universidad Nacional Aut\'onoma de M\'exico,
Circuito Exterior C.U., A.P. 70-543, M\'exico D.F. 04510, M\'exico}

\author{Olivier Sarbach}
\affiliation{Instituto de F\'isica y Matem\'aticas, Universidad Michoacana de San Nicol\'as de Hidalgo,
Edificio C-3, Ciudad Universitaria, 58040 Morelia, Michoac\'an, M\'exico}


\date{\today}


\begin{abstract}
In previous work we constructed new boson star solutions consisting of a family of massive complex scalar fields minimally coupled to gravity in which the individual fields have angular momentum, yet the configuration as a whole is static and spherically symmetric. In the present article we study the linear stability of these $\ell$-boson stars with respect to time-dependent, radial perturbations. The pulsation equations, governing the dynamics of such perturbations are derived, generalizing previous work initiated by M. Gleiser, and shown to give rise to a two-channel Schr\"odinger operator. Using standard tools from the literature, we show that for each fixed value $\ell$ of the angular momentum number, there exists a family of $\ell$-boson stars which are linearly stable with respect to radial fluctuations; in this case the perturbations oscillate in time with given characteristic frequencies which are computed and compared with the results from a nonlinear numerical simulation. Further, there is also a family of $\ell$-boson stars which are linearly unstable. The two families are separated by the configuration with maximum mass. These results are qualitatively similar to the corresponding stability results of the standard boson stars with $\ell=0$, and they imply the existence of new stable configurations that are more massive and compact than usual boson stars.
\end{abstract}


\pacs{
04.20.-q, 
04.25.Dm, 
95.30.Sf, 
98.80.Jk  
}


\maketitle


\section{Introduction}
\label{Sec:Intro}

A boson star~\cite{Kaup68,Ruffini69,Jetzer92,Schunck:2003kk,Liebling:2012fv} is a hypothetical object described by a classical solution to the stationary Einstein-Klein-Gordon (EKG) system which is sourced by a complex, massive scalar field whose time-dependency is harmonic. In their most simple realization boson stars are static and spherically symmetric, although rotating generalizations which are stationary and axisymmetric have also been found long time ago~\cite{1996rscc.conf..138S,Yoshida:1997qf}. Beyond their simplicity, it has recently been shown using numerical evolutions of the fully nonlinear dynamical equations that static and spherically symmetric boson stars also naturally arise as the final state of binary boson star collisions, while rotating boson stars have not been observed to form in such a process~\cite{Palenzuela:2017kcg}. This is due to gravitational cooling \cite{Seidel:1993zk,Sanchis-Gual:2019ljs}, which radiates all the angular momentum of the system.

In a previous paper~\cite{Alcubierre:2018ahf} we showed that boson stars, as they were originally introduced in the late sixties, do not constitute the most general solution to the static spherically symmetric EKG system. Standard boson stars can be easily generalized if the internal symmetry group is extended from $U(1)$ to $U(N)$ with arbitrary odd values of $N$. In this way, the internal group can hide not only the time dependency of the field 
but also the angular dependency of nontrivial harmonics if their amplitudes are excited in an appropriate way~\cite{Olabarrieta:2007di}. We dubbed these new states $\ell$-boson stars in~\cite{Alcubierre:2018ahf}, with $\ell=(N-1)/2$ an arbitrary nonnegative integer.

Astrophysical realizations of $\ell$-boson stars demand such configurations to be dynamically stable. Thus, the study of their stability is of utmost importance. Previous stability studies of boson stars, which correspond to the particular case of $\ell$-boson stars with $\ell=0$ and $N=1$, have been performed based on different approaches. Early studies focused on semi-analytic methods based on linear perturbation theory~\cite{Gleiser:1988rq,Gleiser:1989a,Lee89} in which the EKG system is truncated at linear order. Later, these studies were complemented with nonlinear stability analyses evolving the full EKG equations with the help of numerical codes~\cite{Balakrishna:1997ej,Seidel90,Hawley2000,Guzman09} (see also~\cite{Kusmartsev:1990cr} for an analysis based on catastrophe theory). The results of both types of approaches show that static, spherically symmetric $\ell$-boson stars with $\ell=0$ possess both stable and unstable branches in the solution space, similar to what occurs in spherical relativistic fluid stars~\cite{Chandrasekhar:1964zza, Chandrasekhar:1964zz,Shapiro:1983du}. 

$\ell$-boson stars have $N = 2\ell+1$ scalar fields, and in principle each of these fields can be perturbed in an independent way. However, a simplification occurs if one assumes that the $2\ell+1$ fields are described by the same radial perturbation, in which case spherical symmetry is preserved at the level of the perturbed configurations.
Based on this assumption, in recent work~\cite{Alcubierre:2019qnh} we have performed full nonlinear numerical simulations of the spherical EKG system. Our study indicates that $\ell$-boson stars have similar stability properties than the standard $\ell=0$ stars, namely the $\ell$-boson stars possess both stable and unstable branches of solutions. Moreover, it was found in~\cite{Alcubierre:2019qnh} that small perturbations of stable configurations may exhibit extremely long-lived oscillations.
The main purpose of the present work is to perform a linear stability analysis of $\ell$-boson stars, based on the aforementioned assumption of spherical symmetry, and prove that there is, indeed, a stable branch which is characterized by the absence of growing modes and the presence of oscillatory modes in the spherically symmetric sector.

Regarding the stability of $\ell$-boson stars with respect to nonspherical perturbations, recent numerical evolutions of the full 3D EKG equations indicate that within the timescales explored, the configurations belonging to the spherical stable branch do not possess nonspherical growing modes~\cite{Jaramillo:2020rsv}. In a recent work, Sanchis-Gual, \emph{et.al.}~\cite{Sanchis-Gual:2021edp} have provided numerical evidence that $\ell$-boson stars 
are in fact symmetry-enhanced stable points of larger continuous families of multi-field, multi-frequency boson stars. Although it would be desirable to confirm these findings through a semi-analytic linear stability analysis including nonspherical modes, such a study lies beyond the scope of the present work. Therefore, as mentioned previously, here we restrict ourselves to spherical linear perturbations.

This work is organized as follows. We start in section~\ref{Sec:Spherical} with a brief review on the spherically symmetric field equations and conserved quantities which are relevant for this work and provide a short description of the main behavior of the equilibrium configurations constructed in~\cite{Alcubierre:2018ahf}. Next, in section~\ref{Sec:Pulsation} we derive the pulsation equations, describing the dynamics of linearized radial perturbations of the $\ell$-boson star ground state configurations, generalizing previous work by Gleiser~\cite{Gleiser:1988rq} and by Gleiser and Watkins~\cite{Gleiser:1989a}. These equations have the form of a $2\times 2$ coupled wave system with matrix-valued potential, and for modes with a harmonic time dependency they give rise to a self-adjoint coupled system of radial Schr\"odinger equations. We discuss various properties of the corresponding Schr\"odinger operator, including the asymptotic behavior of the mode solutions near the center and at infinity and the generalized nodal theorem proven in~\cite{hApQ95}, which allows one to determine the number of bound states with negative energy (corresponding to unstable exponentially in time growing modes in our system) by counting the number of zeros of a certain determinant. In Section~\ref{Sec:Results} we present the main results of this paper, starting with a computation for the number of negative energy bound states for different $\ell$-boson star configurations. These results show that, as expected from the numerical simulations in our accompanying paper~\cite{Alcubierre:2019qnh}, the configurations on the stable branch are linearly stable, while the configurations lying on the unstable branch are linearly unstable. Next, based on a shooting algorithm for determining the ground state energy of the Schr\"odinger operator, we compute the oscillation frequencies of the perturbations for the stable configurations, and compare them to the frequencies found in the numerical simulations performed in~\cite{Alcubierre:2019qnh}. Conclusions are drawn in section~\ref{Sec:Conclusions} and technical details used in the numerical integration and shooting algorithm are included in an appendix at the end of the article.

Throughout this work, we use the signature convention $(-,+,+,+)$ for the spacetime metric and Planck units such that $G=c=\hbar=1$. As in our previous articles~\cite{Alcubierre:2018ahf,Alcubierre:2019qnh}, for simplicity, we restrict our attention to the case in which the scalar field is minimally coupled to gravity and is not self-interacting.

\section{Spherically symmetric field equations and equilibrium configurations}
\label{Sec:Spherical}

In this section we review briefly the construction leading to the $\ell$-boson stars~\cite{Alcubierre:2018ahf}. These configurations give rise to a new class of static, spherically symmetric solutions of the EKG system, where the internal global $U(1)$ symmetry of the standard boson star model is promoted to an arbitrary $U(N)$ group. Accordingly, the scalar field $\Phi$ consists of $N$ components, that for convenience can be seen as a collection of $N$ complex scalar fields $\Phi_i$, $i=1,\dots,N$, of equal mass $\mu$ that we choose without self-interaction and minimally coupled to gravity. The spacetime metric describing these solutions (and their time-dependent generalizations) is parametrized in terms of the Misner-Sharp mass $M(t,r)$ and the lapse $\alpha(t,r)$ functions according to
\begin{equation}
ds^2 = -\alpha^2 dt^2 + \gamma^2 dr^2 + r^2 d\Omega^2, \quad  
\gamma^2 := \frac{1}{1 - \frac{2M}{r}} ,
\label{Eq:Metric}
\end{equation}
where $r$ is the areal radial coordinate and $d\Omega^2$ denotes the standard line element on the unit two-sphere. The most simple realization of a nontrivial $\ell$-boson star 
appears for an odd $N$ different from one, with the components of the scalar field given by
\begin{equation}
\Phi_{\ell m}(t,r,\vartheta,\varphi) = \phi_\ell(t,r) Y^{\ell m}(\vartheta,\varphi).
\label{Eq:Ylm}
\end{equation}
Notice that the total angular momentum number $\ell$ is given in terms of a fixed nonnegative integer, and $m$ takes values $m = -\ell,-\ell+1,\ldots,\ell$, for a total of $N=2\ell+1$
components. 
As usual $Y^{\ell m}$ denotes the standard spherical harmonics normalized such that $\sum_{m=-\ell}^\ell |Y^{\ell m}|^2 = (2\ell+1)/4\pi$, and the field amplitude 
$\phi_\ell(t,r)$ is the {\it same} for all $m$. 
As shown in~\cite{Alcubierre:2018ahf} (see also~\cite{Olabarrieta:2007di}), this leads to a total stress energy-momentum tensor\footnote{Notice that in this paper we follow the same conventions as in reference~\cite{Alcubierre:2018ahf} regarding the normalization of the scalar field; hence relative to reference~\cite{Alcubierre:2019qnh} the normalization of the stress energy-momentum tensor, equation~(\ref{eq:EMT}), and the conserved current, equation~(\ref{eq:current}), differs by a factor of $(2\ell+1)/4\pi$.} 
\begin{equation}\label{eq:EMT}
T_{\mu\nu} = \frac{1}{2}\sum_{m = -\ell}^\ell\left[\nabla_\mu\Phi_{\ell m}^*\nabla_\nu\Phi_{\ell m} + \nabla_\mu\Phi_{\ell m}\nabla_\nu\Phi_{\ell m}^*
- g_{\mu\nu}\left( \nabla_\alpha\Phi_{\ell m}^*\nabla^\alpha\Phi_{\ell m} + \mu^2\Phi_{\ell m}^*\Phi_{\ell m} \right)\right],
\end{equation}
($\Phi_{\ell m}^*$ denoting the complex conjugate of $\Phi_{\ell m}$) which is spherically symmetric. If the symmetry group is large enough, different values of $\ell$ could be excited at the same time in the configuration; however, for the purposes of this paper we restrict our attention to the case with only one $\ell$.

With the ans\"atze described in equations~(\ref{Eq:Metric},\ref{Eq:Ylm}), 
the EKG system reduces to the following system of equations~\cite{Alcubierre:2018ahf}:
\begin{subequations}\label{eq:ESS}
\begin{eqnarray}
\dot{M} &=& \kappa_\ell r^2\frac{\alpha}{\gamma} \re( \Pi_\ell^*\chi_\ell ),
\label{Eq:D0m}\\
M' &=& \frac{\kappa_\ell r^2}{2}
\left[  |\Pi_\ell|^2 + |\chi_\ell|^2 + \left( \mu^2 + \frac{\ell(\ell+1)}{r^2} \right) |\phi_\ell|^2 \right],
\label{Eq:mprime}\\
\frac{\alpha'}{\alpha} &=& \gamma^2 \left\{
  \frac{M}{r^2} + \frac{\kappa_\ell r}{2} \left[ |\Pi_\ell|^2 + |\chi_\ell|^2 - \left( \mu^2 + \frac{\ell(\ell+1)}{r^2} \right) |\phi_\ell|^2 \right] \right\},
\label{Eq:D0nu}\\
\dot{\phi}_\ell &=& \alpha\Pi_\ell,
\label{Eq:D0phi}\\
\dot{\Pi}_\ell &=& \frac{1}{r^2\gamma}\left( r^2\frac{\alpha}{\gamma} \phi_\ell' \right)' 
 - \kappa_\ell r \alpha\gamma \re( \Pi_\ell^*\chi_\ell )\Pi_\ell
 - \alpha\left(  \mu^2 + \frac{\ell(\ell+1)}{r^2} \right) \phi_\ell,
\label{Eq:D0Pi}
\end{eqnarray}
\end{subequations}
with $\kappa_\ell = 2\ell+1$ and where $\Pi_\ell := \alpha^{-1}\dot{\phi}_\ell$ and $\chi_\ell := \gamma^{-1}\phi_\ell'$. 
Here and in the following, a dot and a prime denote partial derivatives with respect to $t$ and $r$, respectively. 

\begin{figure}
\includegraphics[width=0.49\textwidth]{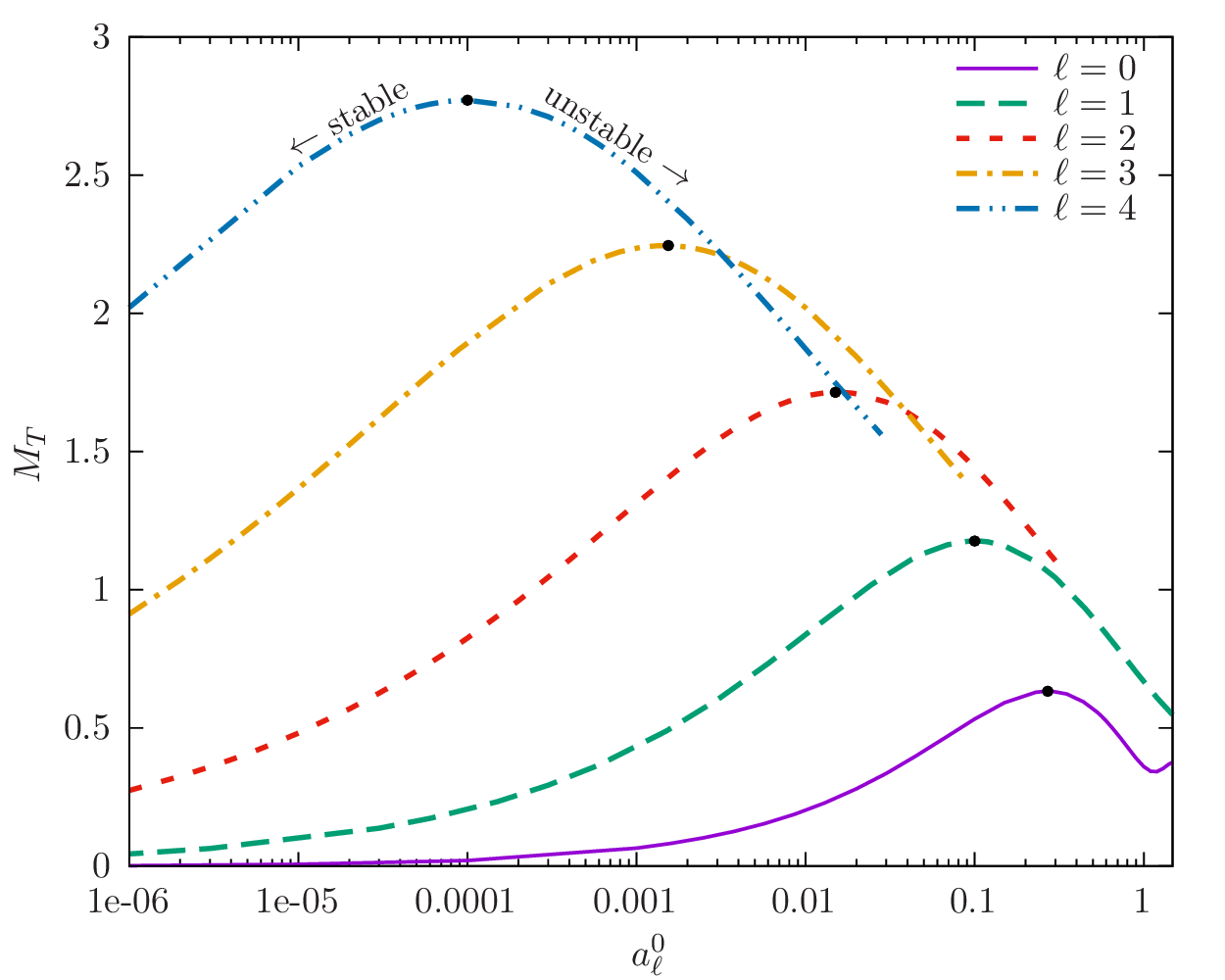}
\includegraphics[width=0.49\textwidth]{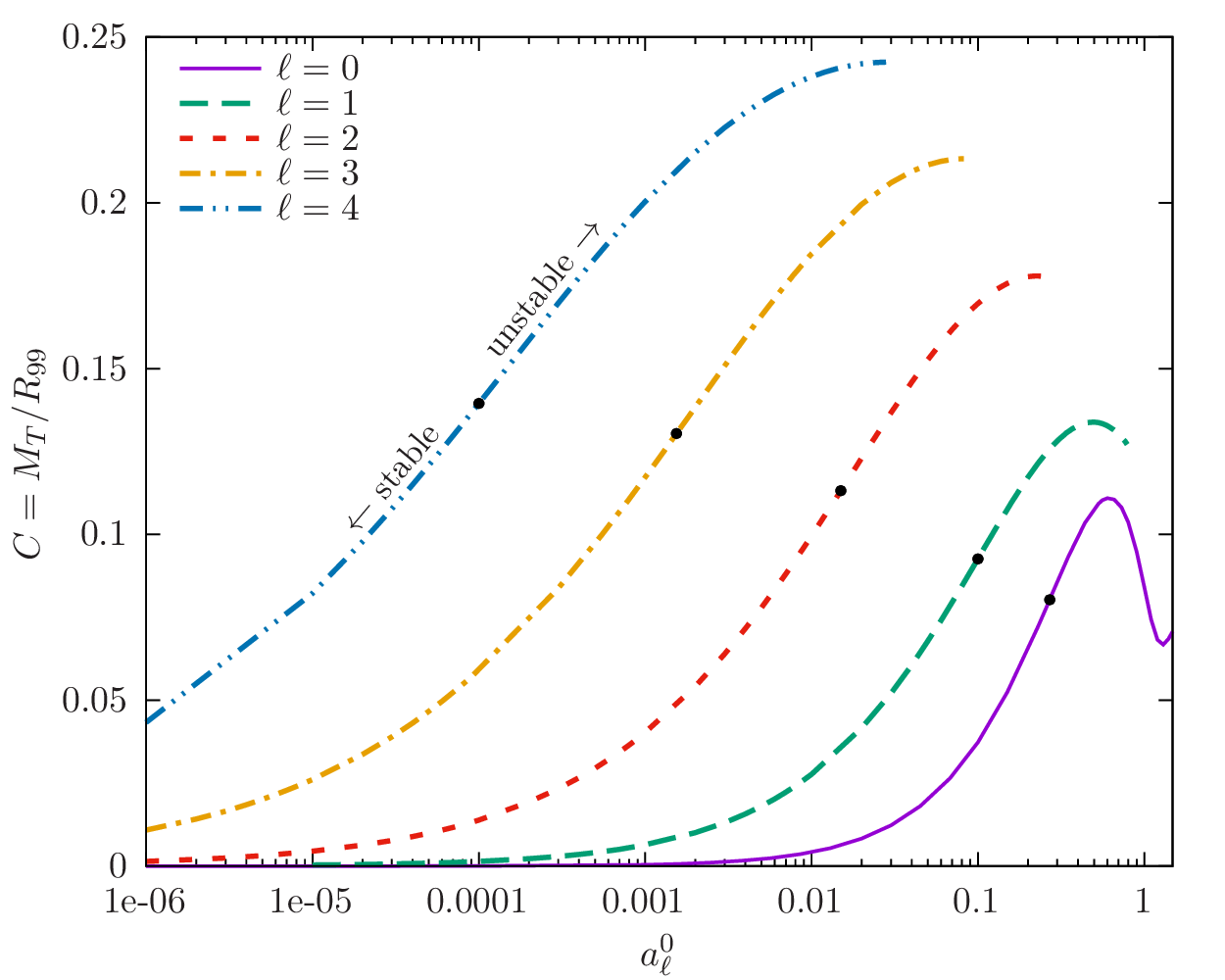}
\caption{The total mass $M_T$ and compacticity $C$ as a function of
$a_\ell^0$
for equilibrium configurations of different angular momentum number $\ell$. Note that the first maximum of $M_T$ divides the stable from the unstable branch. As the value of $\ell$ increases, more massive and compact stable objects are allowed.}
 \label{fig:MvsR}
\end{figure}

The $\ell$-boson star solutions described in~\cite{Alcubierre:2018ahf} are obtained by integrating these equations for the time-harmonic ansatz
\begin{equation}
\phi_\ell(t,r) = e^{i\omega t}\psi_\ell(r),
\label{Eq:TimeHarmonicAnsatz}
\end{equation}
with $\omega$ a real frequency and $\psi_\ell$ a real-valued function of $r$, which behaves as $\psi_\ell\simeq r^\ell$ in the vicinity of the center $r = 0$ and decays exponentially fast as $r\to \infty$. In this way one finds, numerically and for each $\ell$, families of solutions $(\omega,\alpha(r),\gamma(r),\psi_\ell(r))$ which can be parametrized by   $a_\ell^0:=(2\ell+1)^{-1}r^{-\ell}\psi_\ell(r)|_{r=0}$ and the number of nodes of the function $\psi_\ell(r)$ inside the interval $0 < r < \infty$. In this work we restrict our attention to ground state solutions, for which $\psi_\ell$ has no nodes. Similar to the case of the standard boson stars with $\ell=0$, as the value of $a_\ell^0$ increases from zero, the total mass 
\begin{equation}
M_T = \lim_{r\to \infty} M(t,r)
\end{equation}
of the configurations (which coincides with the Arnowitt-Deser-Misner mass) starts increasing, but at some point develops a maximum after which it decreases, see the left panel of figure~\ref{fig:MvsR} for details. The numerical simulations performed in~\cite{Alcubierre:2019qnh} indicate that the configurations belonging to values of $a_\ell^0$ below the one corresponding to the maximum of the mass are stable with respect to small (but nonlinear) spherical perturbations, while the configurations with larger $a_\ell^0$ are unstable, and either collapse to a black hole, or (depending on the sign of the binding energy) migrate to a stable configuration or disperse to infinity. Thus the behavior is analogous to the standard $N=1$ boson star solutions in which the maximum mass configuration divides the solution curve into ``stable" and ``unstable" branches. Similarly, the compacticity, defined as $C = M_T/R_{99}$, where $R_{99}$ is the radius of a sphere containing 99\% of the total mass, starts increasing as $a_\ell^0$ increases, until it develops a maximum. However, note that this maximum occurs at a higher value of $a_\ell^0$ than the one corresponding to the maximum mass configuration, see the right panel of figure~\ref{fig:MvsR}. As we can appreciate from this figure, larger values of $\ell$ allow not only for more massive (stable) solutions, but also for larger compacticities. That is, there exist {\em stable} solutions that can be more compact than the standard $\ell=0$ boson stars. Furthermore, the larger the value of $\ell$, the more compact the stable solutions can be, at least for the first values of the angular momentum number that we have explored in this paper. As mentioned in the introduction, the main goal of this work is to prove that configurations belonging to the ``stable branch" are in fact linearly stable, whereas the remaining ones are linearly unstable.

Like in our accompanying work~\cite{Alcubierre:2019qnh}, the total mass $M_T$ and the total boson number $N_B$, which we define next,
will play an important role in the analysis that we present below. Since the theory is invariant under internal $U(N)$ transformations, there are $N^2$ conserved current densities, one associated with each generator of the internal symmetry group. Among them, there is one corresponding to the total number of particles minus the antiparticles, given by
\begin{equation}\label{eq:current}
J^{\mu} = \frac{i}{2}\sum_{m=-\ell}^\ell\left[
 \Phi^{*}_{\ell m}\nabla^{\mu}\Phi_{\ell m} - \Phi_{\ell m}\nabla^{\mu}\Phi^{*}_{\ell m} \right],
\end{equation}
which is conserved, $\nabla_{\mu}J^{\mu}=0$. For the spherically symmetric configurations~(\ref{Eq:Metric},\ref{Eq:Ylm}) analyzed in this work, $J^\mu$ has vanishing angular components and gives rise to the conserved boson number
\begin{equation}
N_B = (2\ell+1)\int\limits_0^\infty \im[ \phi_\ell^*\Pi_\ell ] \gamma r^2 dr
 = (2\ell +1)\omega\int\limits_0^\infty \psi_\ell^2 \frac{\gamma}{\alpha} r^2 dr,
\label{Eq:ParticleNumber}
\end{equation}
where the second integral is restricted to the time-harmonic ansatz~(\ref{Eq:TimeHarmonicAnsatz}).


\section{Linear stability analysis}
\label{Sec:Pulsation}

After having reviewed the most relevant equations of reference~\cite{Alcubierre:2018ahf}, in this section we perform a linear stability analysis of the $\ell$-boson star ground state configurations. We start in section~\ref{sect.pulsation} by generalizing previous work by Gleiser~\cite{Gleiser:1988rq} and by Gleiser and Watkins~\cite{Gleiser:1989a} to arbitrary values of $\ell$, and in this way we derive a coupled $2\times 2$ wave-like system (the pulsation equations), governing the dynamics of linearized perturbations. By analyzing mode solutions with a harmonic time-dependency, this system reduces to a time-independent system of Schr\"odinger equations, and in section~\ref{Sec:nodal} we establish several important properties of the corresponding Schr\"odinger operator. In particular, we show that  it is (formally) self-adjoint and discuss some tools that we shall use in this article, such as the generalized nodal theorem~\cite{hApQ95}. Sections~\ref{sect.standard.form}, \ref{SubSec:Asym0} and \ref{Sec:asymptotic} are dedicated to a further analysis of the Schr\"odinger operator as well as to the asymptotic behavior of the mode solutions in the vicinity of $r=0$ and as $r\to\infty$. The findings of this section lay the theoretical ground for the numerical results presented in  section~\ref{Sec:Results}.

\subsection{Pulsation equations}\label{sect.pulsation}

In order to derive the pulsation equations it is convenient to write the field amplitude $\phi_\ell(t,r)$ as
\begin{equation}\label{eq.phi_ell}
\phi_\ell(t,r) = 
e^{i\omega t}\left[\psi_{\ell1}(t,r)+i\psi_{\ell2}(t,r)\right],
\end{equation}
where $\psi_{\ell1}(t,r)=\psi_{\ell 0}(r)+ \delta\psi_{\ell 1}(t,r)$ and $\psi_{\ell2}(t,r)= \delta\psi_{\ell 2}(t,r)$ are real-valued, and where $\psi_{\ell 0}(r)$ and $\omega$ are the radial function and frequency, respectively, associated with the background solution. Following~\cite{Gleiser:1989a}, the linearized fields are expanded in the form
\begin{subequations}
\begin{eqnarray}
\delta\psi_{\ell 1}(t,r) &=& \psi_{\ell 0}(r)\delta\varphi_{\ell1}(t,r),
\label{Eq:deltapsiell1}\\
\delta\psi_{\ell 2}(t,r) &=& \psi_{\ell 0}(r)\delta\varphi_{\ell2}(t,r),
\label{Eq:deltapsiell2}\\
\delta\alpha(t,r) &=& \frac{1}{2}\alpha_0(r)\delta\nu(t,r),\\
\delta\gamma(t,r) &=& \frac{1}{2}\gamma_0(r)\delta\lambda(t,r),
\end{eqnarray}
\end{subequations}
giving rise to small (but time-dependent) variations of the static configurations $(\omega,\alpha_0(r),\gamma_0(r),\psi_{\ell 0}(r))$ introduced in the previous section. Here the quantity $\delta\lambda$ is related to the linearized mass function $\delta M$ via the relation 
$\delta\lambda = 2\gamma_0^2\delta M/r$. Note that there is a gauge ambiguity in the definition of the perturbations $\delta\varphi_{\ell2}$ and $\delta\nu$, since one can still perform a redefinition of the time coordinate $t \mapsto \tilde{t}=\tilde{t}(t)$ and maintain the same form of the line element as in equation~(\ref{Eq:Metric}). Under an infinitesimal transformation $t \mapsto \tilde{t} = t+f(t)$, where $f(t)$ is an arbitrary function of time, the fields $\delta\varphi_{\ell1}(t,r)$ and $\delta\lambda(t,r)$ remain unaltered, whereas $\delta\varphi_{\ell2}(t,r)$ and $\delta\nu(t,r)$ change according to:
\begin{subequations}\label{eq.gauge}
\begin{eqnarray}
\delta\varphi_{\ell2}(t,r) &\mapsto& \delta\tilde{\varphi}_{\ell2}(t,r)=\delta\varphi_{\ell2}(t,r)-\omega f(t),\\
\delta\nu(t,r) &\mapsto& \delta\tilde{\nu}(t,r)=\delta\nu(t,r) -2\dot{f}(t).
\end{eqnarray}
\end{subequations}
We stress that the decomposition~(\ref{Eq:deltapsiell1},\ref{Eq:deltapsiell2}) is only valid for the ground state configurations for which $\psi_\ell(r)$ has no zeros in the interval $0 < r < \infty$. Further below we shall also assume that the linearized fields $\delta\varphi_{\ell1}$, $\delta\varphi_{\ell2}$, $\delta\alpha$ and $\delta\lambda$ have a harmonic time-dependency of the form $e^{-i\sigma t}$.

Linearizing equations~(\ref{Eq:D0m},\ref{Eq:mprime},\ref{Eq:D0nu}), one obtains the following useful relations between the linearized metric 
coefficients and scalar fields:
\begin{subequations}\label{Eq.constraints}
\begin{eqnarray}
\delta\dot{M} &=& \kappa_\ell\frac{r^2\psi_\ell}{\gamma^2}
\left[ \psi_\ell'\delta\dot{\varphi}_{\ell1} + \omega\psi_\ell\delta\varphi_{\ell2}' \right],
\label{Eq:dmdot}\\
\delta M' &=& \kappa_\ell r^2\left\{ \left[ \frac{\psi_\ell'^2}{\gamma^2} + \left( \mu^2 + \frac{\ell(\ell+1)}{r^2} + \frac{\omega^2}{\alpha^2} \right) \psi_\ell^2 \right]\delta\varphi_{\ell1} 
 + \frac{\psi_\ell\psi_\ell'}{\gamma^2}\delta\varphi_{\ell1}'  
 + \frac{\omega\psi_\ell^2}{\alpha^2}\delta\dot{\varphi}_{\ell2}
 - \frac{\omega^2\psi_\ell^2}{2\alpha^2}\delta\nu - \frac{\psi_\ell'^2}{2\gamma^2}\delta\lambda \right\},
\label{Eq:dmprime}\\
\frac{\delta\nu' - \delta\lambda'}{4} &=& \frac{\gamma^4}{r^2}\delta M
 - \kappa_\ell\gamma^2\psi_\ell^2\left( \mu^2 +\frac{\ell(\ell+1)}{r^2} \right)
 (r\delta\varphi_{\ell1} + \gamma^2\delta M),
\label{Eq:dnu}
\end{eqnarray}
\end{subequations}
where from now on we omit the subscript $0$ on the background quantities to simplify the notation. Eliminating $\delta\varphi_{\ell2}$ from the first two equations and linearizing the remaining equations~(\ref{Eq:D0phi},\ref{Eq:D0Pi}), one obtains the following evolution system:
\begin{subequations}\label{eq:original.system}
\begin{eqnarray}
 \delta\varphi_{\ell 1}'' + \left[\frac{2}{r} + \frac{\alpha'}{\alpha} - \frac{\gamma'}{\gamma} \right]\delta\varphi_{\ell 1}' + \frac{1}{\kappa_\ell r\psi_\ell^2}\delta\lambda' 
 - \frac{\gamma^2}{\alpha^2}\delta\ddot{\varphi}_{\ell 1}&& \nonumber\\
 +\left\lbrace \frac{1 - 2r\frac{\gamma'}{\gamma}}{\kappa_{\ell}r^2\psi_\ell^2} + \frac{\psi_\ell'}{\psi_\ell}\left[\frac{\alpha'}{\alpha} - \frac{\gamma'}{\gamma} + \frac{\psi_\ell'}{\psi_\ell} + \frac{1}{r} \right]
- \gamma^2\left[\mu^2+\frac{\ell(\ell+1)}{r^2} - \frac{\omega^2}{\alpha^2} \right] \right\rbrace \delta\lambda &&\nonumber\\
 -2\gamma^2\left\lbrace \mu^2+\frac{\ell(\ell+1)}{r^2} +  \frac{\omega^2}{\alpha^2}
  + \frac{\psi_\ell'^2}{\gamma^2\psi_\ell^2} + \kappa_\ell r\left[\mu^2+\frac{\ell(\ell+1)}{r^2}\right]\psi_\ell\psi_\ell'\right\rbrace \delta\varphi_{\ell 1}&=&0,
\label{Eq:deltaphi}
\end{eqnarray}
\begin{eqnarray}
 \delta\lambda'' + 3\left( \frac{\alpha'}{\alpha} - \frac{\gamma'}{\gamma} \right)\delta\lambda' + 4\kappa_\ell\left\lbrace 2\psi_\ell\psi'_\ell - r\gamma^2\left[\mu^2+\frac{\ell(\ell+1)}{r^2}\right]\psi_\ell^2 \right\rbrace \delta\varphi_{\ell1}' - \frac{\gamma^2}{\alpha^2}\delta\ddot{\lambda}\nonumber\\
 - 2\left\lbrace 2\kappa_{\ell}\psi'^2_{\ell} + \frac{1}{r^2} + \left( \frac{\gamma'}{\gamma} \right)'
 - \left( \frac{\alpha'}{\alpha} - \frac{\gamma'}{\gamma} \right)^2 
 - \frac{1}{r}\left( 2\frac{\alpha'}{\alpha} + \frac{\gamma'}{\gamma} \right) \right\rbrace \delta\lambda 
&& \nonumber\\
 +4\kappa_\ell\left\lbrace 2\psi_\ell'^2 - r\gamma^2\left[\mu^2+\frac{\ell(\ell+1)}{r^2}\right]\psi_\ell^2\left[ 2\frac{\psi'_{\ell}}{\psi_\ell} + 2\frac{\alpha'}{\alpha} + \frac{\gamma'}{\gamma} \right] + \gamma^2\frac{\ell(\ell+1)}{r^2}\psi_\ell^2 \right\rbrace \delta\varphi_{\ell1} &=& 0,
\label{Eq:deltalambda}
\end{eqnarray}
\end{subequations}
which reduce to equations~(34) and (35) in~\cite{Gleiser:1989a} when $\ell=0$ (taking into account a different sign convention for the frequency $\omega$ and a factor $e^\lambda$ which was corrected  in~\cite{Hawley2000}). We refer to this system as the pulsation equations in this paper. 
Note that it relies only on quantities that are invariant under infinitesimal time redefinitions $t \mapsto \tilde{t} = t+f(t)$.

To solve equations~(\ref{eq:original.system}) we need to specify suitable initial data. In order to do so we fix the value of the perturbations of the field $\phi_{\ell}(t,r)$ defined in equation~(\ref{eq.phi_ell}) and its conjugate momentum, $\Pi_\ell(t,r)=\alpha^{-1}\dot{\phi}_\ell$, at time $t=0$,
\begin{subequations}
\begin{eqnarray}
\delta\phi_\ell(0,r) &=& \psi_\ell(r)\left[ \delta\varphi_{\ell 1}(0,r) + i\delta\varphi_{\ell 2}(0,r) \right],\\
\delta\Pi_\ell(0,r) &=&  
\frac{\psi_\ell(r)}{\alpha(r)}\left[
\delta\dot{\varphi}_{\ell 1}(0,r) - \omega\delta\varphi_{\ell 2}(0,r) + i\left( \delta\dot{\varphi}_{\ell 2}(0,r) + \omega\delta\varphi_{\ell 1}(0,r) - \frac{\omega}{2}\delta\nu(0,r) \right) \right].
\end{eqnarray}
\end{subequations}
This determines univocally the initial data for the pulsation equations: 
\begin{subequations}
\begin{eqnarray}
\delta\varphi_{\ell1}(0,r) &=& \frac{1}{\psi_{\ell}(r)}\re\delta\phi_\ell(0,r),\\
\delta\dot{\varphi}_{\ell1}(0,r) &=& \frac{\alpha(r)}{\psi_{\ell}(r)}\re\delta\Pi_\ell(0,r)+\frac{\omega}{\psi_\ell(r)}\im\delta\phi_\ell(0,r),\\
\delta M(0,r) &=& \kappa_\ell\int\limits_0^r e^{a(s,r)}
\left[ \frac{\omega}{\alpha(s)}\im\delta\Pi_\ell(0,s) + \left( \mu^2 + \frac{\ell(\ell+1)}{s^2} \right)\re\delta\phi_\ell(0,s) +   \frac{\psi_\ell'(s)}{\psi_\ell(s)}\frac{\re\delta\phi_\ell'(0,s)}{\gamma^2(s)}\right]\psi_\ell(s) s^2 ds,\\
\delta\dot{M}(0,r) &=& \kappa_\ell\frac{r^2\psi_\ell(r)}{\gamma^2(r)}\left[
\frac{\psi_{\ell}'(r)}{\psi_\ell(r)}\alpha(r)\re\delta\Pi_\ell(0,r)
+\frac{\psi_{\ell}'(r)}{\psi_\ell(r)}\omega\im\delta\phi_\ell(0,r
)
+\omega\im\delta\phi_\ell'(0,r)
\right],
\end{eqnarray}
\end{subequations}
where
\begin{equation}
a(r_1,r_2):=-\kappa_\ell\int\limits_{r_1}^{r_2} r\psi_\ell'^2(r) dr.
\end{equation}
(Remember that $\delta M$ is related to $\delta\lambda$ through a background function.)
In order to obtain $\delta M(0,r)$ we have integrated equation~(\ref{Eq:dmprime}) under the assumption that $\delta M(0,0)=0$, a condition that is necessary to guarantee regularity at the origin, and $\delta\dot{M}(0,r)$ was obtained from the constraint~(\ref{Eq:dmdot}).
Once these equations are solved, one acquires $\delta\nu$ by integrating equation~(\ref{Eq:dnu}), and subsequently $\delta\varphi_{\ell2}$ is obtained by integrating equations~(\ref{Eq:dmdot},\ref{Eq:dmprime}).
Note that these two last quantities are defined only up to the gauge ambiguity that we identified in~(\ref{eq.gauge}).

For practical purposes we will restrict our attention to initial data $\delta\phi_\ell(0,r)$ and $\delta\Pi_\ell(0,r)$ that are smooth and of compact support in the interval $(0,\infty)$. Furthermore, we shall require that the initial data satisfies $\delta M_T := \lim_{r\to\infty} \delta M(0,r) = 0$, such that the perturbation does not change the total mass of the system. These conditions imply that $\delta M(0,r)$ and $\delta\lambda(0,r)$ are also compactly supported on $(0,\infty)$ and that the total particle number is unaffected. In order to see this more explicitly note that linearizing equation~(\ref{Eq:ParticleNumber}) one obtains [cf. equation~(37) in~\cite{Gleiser:1988rq}]
\begin{equation}
\delta N_B = (2\ell + 1)\int\limits_0^\infty \left[ \frac{\omega}{2}(\delta\lambda - \delta\nu) 
 + \delta\dot{\varphi}_{\ell2} + 2\omega\delta\varphi_{\ell1} \right] \frac{\gamma}{\alpha} r^2\psi_\ell^2 dr.
\label{Eq:deltaN}
\end{equation}
After some manipulations using the constraints~(\ref{Eq.constraints}) and the background equations~(\ref{eq:ESS}) the integrand in the last expression can be re-written as a total differential, and one obtains the simpler expression
\begin{equation}
\delta N_B(t) = \left. 
(2\ell +1)\left[\frac{\alpha(r)}{\omega\gamma(r)} \left(
\frac{\gamma^2(r)}{\kappa_\ell}\delta M(t,r) - r^2\psi_{\ell}'(r)\psi_{\ell}(r)\delta\varphi_{\ell1}(t,r)\right)\right]\right|_{r=0}^\infty.
\label{Eq.perturbation.number}
\end{equation}
This equation has various important implications. First, it follows that any perturbation for which $\re\left(\delta\phi_\ell(t,r)\right) = \psi_\ell(r)\delta\varphi_{\ell1}(t,r)$ is bounded at $r=0$ and near $r\to\infty$ yields the relation
\begin{equation}
\delta N_B = \frac{1}{\omega }\delta M_T
\label{Eq:NBMTRelation}
\end{equation}
between the total particle number and total mass. In particular, it follows for such perturbations that $\delta N_B = 0$ if and only if $\delta M_T = 0$, and hence the class of compactly supported initial data described above leaves both the total particle number and mass invariant (to linear order in the perturbation). A further consequence of equation~(\ref{Eq:NBMTRelation}) (when applied to the solution curves of static $\ell$-boson stars) is that the total particle number $N_B$ and mass $M_T$ as functions of $a_\ell^0$ have the same critical points, and this explains why the location of the extrema of $N_B$ and $M_T$ coincide (see figures~1 in~\cite{Gleiser:1988rq,Gleiser:1989a} for the $\ell=0$ case, and figure~1 in~\cite{Alcubierre:2019qnh} for $\ell$-boson stars with $\ell=1$). Another consequence of equation~(\ref{Eq:NBMTRelation}) is that the first variation of the binding energy
$E_B := M_T - \mu N_B$
satisfies $\delta E_B = -(\mu-\omega)\delta N_B$. Since $\mu-\omega >0$ this implies that a maximum of $N_B$ corresponds to a minimum of $E_B$ and the other way around. Finally, and most importantly for the purpose of this article, the existence of static perturbations at such extrema signals (but does not prove) the existence of mode solutions to the linearized equations which transition from stable to unstable, see the discussions in section~4 of reference~\cite{Gleiser:1989a} and in section~\ref{SubSec:Modes} below.

\subsection{Spectral properties and nodal theorem}\label{Sec:nodal}

An alternative way of writing the pulsation equations is based on the original work in~\cite{Gleiser:1988rq} where the new quantities $f_1$ and $f_2$ are introduced, which are related to $\delta\varphi_{\ell1}$ and $\delta\lambda$ through the expressions\footnote{Note that $\dot{f}_2 = \delta\varphi_{\ell2}'$ according to equation~(\ref{Eq:dmdot}). Also the perturbation of the boson number, equation~(\ref{Eq.perturbation.number}), can be expressed in the more compact form [cf. equation~(38) in~\cite{Gleiser:1988rq}] $\delta N_B = (2\ell+1)\frac{\alpha}{\gamma}r^2\psi_{\ell}^2 f_2 |_{r=0}^\infty$.}
\begin{equation}
f_1 = \delta\varphi_{\ell1},\qquad
f_2 = \frac{1}{\omega}\left[ \frac{\delta\lambda}{2\kappa_\ell r\psi_\ell^2} 
 - \frac{\psi_\ell'}{\psi_\ell}\delta\varphi_{\ell 1} \right].
\label{Eq:fDef}
\end{equation}
The system for $f := \left( \begin{array}{c} f_1 \\ f_2 \end{array} \right)$ can be written in the form
\begin{equation}
\frac{\gamma^2}{\alpha^2} A\ddot{f} = \frac{d}{dr}\left( A\frac{df}{dr} \right)
 + \frac{d}{dr}(B f) - B^T\frac{df}{dr} + C f,
\label{Eq:fSystem}
\end{equation}
with the matrices $A$, $B$ and $C$ given by
\begin{equation}
A := \frac{\alpha}{\gamma} r^2\psi_\ell^2
\left( \begin{array}{cc} 1 & 0 \\ 0 & \frac{\alpha^2}{\gamma^2} \end{array} \right),\qquad
B := 2\omega\frac{\alpha}{\gamma} r^2\psi_\ell^2
\left( \begin{array}{cc} 0 & 1 \\ 0 & 0 \end{array} \right),
\qquad
C := \frac{\alpha}{\gamma} r^2\psi_\ell^2
\left( \begin{array}{cc} V_{11} & V_{12} \\ V_{21} & V_{22} \end{array} \right),
\end{equation}
and the functions $V_{11}$, $V_{12} = V_{21}$, $V_{22}$ defined by
\begin{subequations}
\begin{eqnarray}
V_{11} &=& -4\gamma^2\frac{\omega^2}{\alpha^2} - 2\kappa_\ell r\psi_\ell'\left[
 \psi_\ell'' + \frac{\psi_\ell'}{r} + \gamma^2\left( \mu^2 + \frac{\ell(\ell+1)}{r^2} + \frac{\omega^2}{\alpha^2} \right)\psi_\ell \right]\nonumber\\
&=&-4\gamma^2\frac{\omega^2}{\alpha^2} - 2\kappa_\ell r\psi_\ell'\left[
2 \gamma^2\left(\mu^2+\frac{\ell(\ell+1)}{r^2}\right)\psi_\ell+\left(\frac{\gamma'}{\gamma}-\frac{\alpha'}{\alpha}-\frac{1}{r}\right)\psi_\ell'\right],
\label{Eq:V11}\\
V_{12} &=& -2\kappa_\ell r\omega\psi_\ell\left( \psi_\ell'' + \frac{\psi_\ell'}{r} + \gamma^2\frac{\omega^2}{\alpha^2}\psi_\ell \right)\nonumber\\
&=& -2\kappa_\ell r\omega\psi_\ell\left[ \gamma^2\left(\mu^2+\frac{\ell(\ell+1)}{r^2}\right)\psi_\ell+\left(\frac{\gamma'}{\gamma}-\frac{\alpha'}{\alpha}-\frac{1}{r}\right)\psi'_\ell\right]\,,
\label{Eq:V12}\\
V_{22} &=& \frac{\alpha^2}{\gamma^2}\left[
 \left( \frac{\alpha'}{\alpha} - \frac{\gamma'}{\gamma} \right)' 
 + 2\left( \frac{\alpha'}{\alpha} - \frac{\gamma'}{\gamma} \right)
 \left( \frac{2}{r} + 2\frac{\psi_\ell'}{\psi_\ell} + \frac{\alpha'}{\alpha} - \frac{\gamma'}{\gamma} \right)
  - \frac{2}{r^2} + 2\frac{\psi_\ell''}{\psi_\ell} - 2\frac{\psi_\ell'^2}{\psi_\ell^2} \right]
  + 2\kappa_\ell\omega^2 r\psi_\ell^2\left( \frac{1}{r}
  + \frac{\alpha'}{\alpha} - \frac{\gamma'}{\gamma} \right)\nonumber\\
  &=&2\alpha^2\frac{\gamma'}{r \gamma}-\frac{\alpha^2(\gamma^2-1)}{r^2}-(2\ell+1)\alpha^2\gamma^2\psi_\ell^2
  \left[\mu^2-\frac{\ell(\ell+1)}{r^2}+r\left(\frac{\gamma'}{\gamma}+\frac{\psi'_\ell}{\psi_\ell}\right)\left(\frac{\ell(\ell+1)}{r^2}+\mu^2\right)\right]\nonumber\\
  &+&\frac{2\alpha^2}{\gamma^2}\left(\frac{\alpha'}{\alpha}-\frac{\gamma'}{\gamma}\right)\left(\frac{\psi'_\ell}{\psi_\ell}+\frac{2}{r}+\frac{\alpha'}{\alpha}\right)-4\frac{\alpha^2}{\gamma^2}\frac{\psi'_\ell}{\psi_\ell}\left(\frac{1}{r}+\frac{\psi'_\ell}{\psi_\ell}\right)-\frac{2\alpha^2}{r^2 \gamma^2}
  +2\alpha^2\left(\mu^2-\frac{\omega^2}{\alpha^2}+\frac{\ell(\ell+1)}{r^2}\right)\,.
\label{Eq:V22}
\end{eqnarray} 
\end{subequations}

Equation~(\ref{Eq:fSystem}) has the form $\ddot{f} = -{\cal H}f$, with ${\cal H}$ the Schr\"odinger-type operator given by
\begin{equation}
{\cal H} = \frac{\alpha^2}{\gamma^2} A^{-1}
\left[ -\frac{d}{dr} A \frac{d}{dr} - \frac{d}{dr} B + B^T\frac{d}{dr} - C \right].\label{Eq:H}
\end{equation}
Due to the fact that the matrix $A = A^T > 0$ is symmetric positive definite and that the matrix $C = C^T$ is symmetric, the operator ${\cal H}$ is formally self-adjoint with respect to the scalar product
\begin{equation}
\langle f, g \rangle := \int\limits_0^\infty f(r)^T A g(r) \frac{\gamma^2}{\alpha^2} dr.
\end{equation}
This means that ${\cal H}$ satisfies
\begin{equation}
\langle f, {\cal H} g \rangle 
 = \langle {\cal H} f, g \rangle
\label{Eq:HSymmetry}
\end{equation}
for all sufficiently smooth functions $f$ and $g$ which are compactly supported on the interval $(0,\infty)$, or, more generally, which satisfy appropriate boundary conditions at $r=0$ and as $r\to\infty$. (We shall analyze these conditions further below.) Since ${\cal H}$ commutes with complex conjugation, it follows from von Neumann's theorem (see Theorem X.3 in~\cite{ReedSimon80II}) that ${\cal H}$ possesses a self-adjoint extension. This offers the possibility of studying the dynamics of the pulsation equation~(\ref{Eq:fSystem}) using the powerful tools of spectral theory for self-adjoint operators, for which there exists a vast literature, see for instance~\cite{ReedSimon80I,ReedSimon80IV}.

The precise determination of the appropriate self-adjoint extension of ${\cal H}$ and its properties lie way beyond the scope of this article. Instead, in what follows, we shall focus on the point spectrum of the operator ${\cal H}$ (i.e. its eigenvalues and eigenfunctions), which in an analogous quantum mechanical problem would correspond to the energy levels $E$ of the bound states. In our scenario, each negative eigenvalue $E < 0$ gives rise to a pair of mode solutions of $\ddot{f} = -{\cal H} f$ proportional to $e^{\pm\sqrt{-E} t}$ which are exponentially growing or decaying in time, and thus a negative eigenvalue implies the instability of the system. On the other hand, each positive eigenvalue $E > 0$ of ${\cal H}$ corresponds to a pair of purely oscillating modes proportional to $e^{\pm i\sqrt{E} t}$.

One of the useful tools we shall apply in the next section is the nodal theorem by Amann and Quittner~\cite{hApQ95}, which allows one to determine the number of negative eigenvalues of ${\cal H}$ by counting the zeros of a certain determinant which is constructed from two linearly independent zero modes of ${\cal H}$. More precisely, one solves the differential system ${\cal H} f^{(j)} = 0$ with initial data $f^{(j)}(r_0) = 0$ and $\frac{d}{dr} f^{(j)}(r_0) = {\bf e}_j$, $j=1,2$, with two linearly independent vectors ${\bf e}_1$ and ${\bf e}_2$ in $\Real^2$. Then, for $r_0 > 0$ small enough and $r_1 > r_0$ large enough, the number of zeros of the determinant function $D(r) := \det(f^{(1)}(r),f^{(2)}(r))$ on the interval $(r_0,r_1)$ is equal to the number of bound states with negative energy (counted with multiplicities) of the operator ${\cal H}$. The number of zeros is independent of $r_0$ and $r_1$, provided $r_0$ is sufficiently close to $0$ and $r_1$ sufficiently large. Moreover, the number of zeros is independent of the choice of the basis vectors ${\bf e}_1$ and ${\bf e}_2$.

In addition to determining the number of negative eigenvalues of ${\cal H}$ (which correspond to the number of unstable, exponentially in time growing mode solutions of the pulsation equations), we shall also compute numerically the eigenvalues $E = \sigma^2$ of ${\cal H}$ by means of a shooting algorithm. Since ${\cal H}$ is formally self-adjoint, $\sigma$ is either real or purely imaginary. In the latter case, the norm of $\sigma$ determines the growth rate of the unstable mode, while in the former case it determines the oscillatory frequency of the mode solutions, which will be compared to the results from a nonlinear numerical time evolution in section~\ref{Sec:Results}.

\subsection{Transformation of the Schr\"odinger operator to a simpler form}
\label{sect.standard.form}

The application of the nodal theorem requires the satisfaction of certain hypotheses we would like to comment on. To this purpose, we first transform the Schr\"odinger operator ${\cal H}$ to a simpler form in which the first-derivative terms are eliminated and the transformed operator is formally self-adjoint with respect to the usual scalar product for square-integrable ($L^2$-) functions. This new form will also simplify the analysis for the asymptotic behavior of the mode solutions in the limits $r\to 0$ and $r\to \infty$.

The transformation we apply is $f = T v$ with
\begin{equation}
T = \sqrt{\frac{\alpha}{\gamma}} \frac{1}{r\psi_\ell}
\left( \begin{array}{cc} 1 & 0 \\ 0 & \frac{\gamma}{\alpha} \end{array} \right) R(r),
\label{Eq:Transform}
\end{equation}
where $R(r)$ is a rotation matrix given by
\begin{equation}
R(r) = \left( \begin{array}{rr} \cos(\omega r_*) & -\sin(\omega r_*) \\ 
\sin(\omega r_*) & \cos(\omega r_*) 
\end{array} \right),\quad
r_* := \int\limits_0^r \frac{\gamma(\bar{r})}{\alpha(\bar{r})} d\bar{r}.
\label{Eq:RDef}
\end{equation}
This transforms the problem $\ddot{f} = -{\cal H} f$ into $\ddot{v} = -{\cal H}_T v$, in which the new operator ${\cal H}_T := T^{-1} {\cal H} T$ has the structurally simpler form
\begin{equation}
{\cal H}_T = -\frac{d}{dr}\left( \frac{\alpha^2}{\gamma^2}\frac{d}{dr} \right) + R^T W R,
\label{Eq:HT}
\end{equation}
with the symmetric matrix $W = W^T$ given by
\begin{equation}
W = -\left( \begin{array}{cc} \frac{\alpha^2}{\gamma^2} V_{11} & \frac{\alpha}{\gamma} V_{12} \\
 \frac{\alpha}{\gamma} V_{21} & V_{22} \end{array} \right)
  + \frac{d}{dr}\left( \frac{\alpha^2}{\gamma^2} Q \right)
  + \frac{\alpha^2}{\gamma^2} Q^2
  - 2\omega\frac{\alpha}{\gamma}\frac{(r\psi_\ell)'}{r\psi_\ell}
  \left( \begin{array}{cc} 0 & 1 \\ 1 & 0 \end{array} \right) - \omega^2 I_2,
\label{Eq:W}
\end{equation}
where
\begin{equation}
Q := \frac{(r\psi_\ell)'}{r\psi_\ell} I_2 
 + \frac{1}{2}\left( \frac{\alpha'}{\alpha} - \frac{\gamma'}{\gamma} \right)
 \left( \begin{array}{rr} -1 & 0 \\ 0 & 1 \end{array} \right),
\label{Eq:Q}
\end{equation}
and $I_2$ denotes the $2\times 2$ identity matrix. Note that the transformed operator ${\cal H}_T$ is formally self-adjoint with respect to the standard $L^2$ scalar product
\begin{equation}
(v,w) := \int\limits_0^\infty v(r)^T w(r) dr.
\end{equation}
More generally, one has the identity
\begin{equation}
(v,{\cal H}_T w) = ({\cal H}_T v,w)
 + \frac{\alpha^2}{\gamma^2}
 \left[ \frac{dv^T}{dr} w - v^T\frac{dw}{dr} \right]_{r=0}^\infty,
\label{Eq:SymId}
\end{equation}
for any pair of twice continuously differentiable functions $v$ and $w$ on $[0,\infty)$, and the boundary term vanishes if $v$ and $w$ are zero at $r=0$ and decay sufficiently rapidly as $r\to \infty$. It is simple to verify that these conditions are automatically satisfied for the type of initial data specified towards the end of section~\ref{sect.pulsation}, and hence the initial data belongs to the class of functions for which the operator ${\cal H}_T$ is self-adjoint.

Next, we note that the function $\alpha^2/\gamma^2$ appearing between the derivative operators $d/dr$ in equation~(\ref{Eq:HT}) is smooth, strictly positive and possesses the limits $\alpha_c^2 := \alpha(0)^2 > 0$ and $1$ as $r\to 0$ and $r\to \infty$, respectively. Furthermore, as follows from the results presented in the next two subsections, the transformed potential $R^T W R$ is smooth on $(0,\infty)$, uniformly bounded near infinity, and near $r=0$ it has the form given in equation~(\ref{Eq:RTWR0}) below with the $1/r^2$ matrix coefficient being nonnegative. The only additional assumptions made by the nodal theorem~\cite{hApQ95} are the requirements that the essential spectrum of ${\cal H}$ contains no negative values and that ${\cal H}$ has only a finite number of negative eigenvalues.\footnote{See~\cite{ReedSimon80I} for a definition of the essential spectrum of an operator and~\cite{ReedSimon80IV} for theorems on its properties and estimates on the number of bound states.} The strict verification of these last two conditions goes beyond the scope of this article; however, the numerical results presented in the next section offer a picture that is fully consistent with the results from the nodal theorem.

In the following, we analyze the asymptotic behavior of the effective potential $R^T W R$ for $r\to 0$ and $r\to \infty$, and the corresponding behavior of the mode solutions.

\subsection{Asymptotic behavior for $r\to 0$}
\label{SubSec:Asym0}

As has been discussed in~\cite{Alcubierre:2018ahf}, the background solution has the following behavior near the center $r=0$:
\begin{subequations}\label{Eqs.behavior.origin}
\begin{eqnarray}
\alpha(r) &=& \alpha_c[ 1 + {\cal O}(r^{2\ell+2})],
\label{Eq:alpha0}\\
\gamma(r) &=& 1 + \frac{1}{2}\ell\kappa_\ell a_0^2 r^{2\ell} + {\cal O}(r^{2\ell+2}),
\label{Eq:gamma0}\\
\psi_\ell(r) &=& a_0 r^\ell[ 1 + a_2 r^2 + {\cal O}(r^4) ],
\label{Eq:psi0}
\end{eqnarray}
\end{subequations}
for some positive constants $\alpha_c$ and $a_0$, with $a_0$ related to the previously defined quantity $a_\ell^0$ through the relation $a_0 =(2\ell+1)a_\ell^0$. The expressions above include the next-to-leading order terms which will be required for the analysis in this subsection and the appendix. They can be computed from the background equations~(17a--17c) in~\cite{Alcubierre:2018ahf}.
In particular, the coefficient $a_2$ in equation~(\ref{Eq:psi0}) can be determined by taking the limit $r\to 0$ of the right-hand side of equation~(17c) in~\cite{Alcubierre:2018ahf}, giving
\begin{equation}
a_2 = \frac{1}{2(2\ell + 3)}\left( \mu^2 - \frac{\omega^2}{\alpha_c^2} 
 + 3\delta_{\ell,1}\kappa_\ell a_0^2 \right).
\label{Eq:a2}
\end{equation}
It follows from equations~(\ref{Eq:alpha0},\ref{Eq:gamma0}) and the definition of $r_*$ in equation~(\ref{Eq:RDef}) that
\begin{equation}
r_* = \frac{r}{\alpha_c}\left[ 1 + \frac{1}{2}\frac{\ell \kappa_\ell}{2\ell + 1}a_0^2 r^{2\ell}
 + {\cal O}(r^{2\ell+2}) \right] = \frac{r}{\alpha_c}\left[ 1 + {\cal O}(r^2) \right],
\label{Eq:r*0}
\end{equation}
for all $\ell\geq 0$.

After these preliminary remarks regarding the properties of the background solution near $r = 0$, we analyze the behavior of the solutions $v$ to the mode equation ${\cal H}_T v = \sigma^2 v$. For the remaining of this subsection, we only compute the leading order terms in $R^T W R$ and $v$. Higher-order contributions (which are required for the numerical shooting algorithm used in the next section) are worked out in the appendix. First, it follows from equations~(\ref{Eq:RDef}) and~(\ref{Eq:r*0}) that
\begin{equation}
R(r) = I_2 + \frac{\omega r}{\alpha_c} \left( \begin{array}{rr} 0 & -1 \\ 1 & 0 \end{array} \right)
+  {\cal O}(r^2).
\end{equation}
Next, from equations~(\ref{Eq:V11}--\ref{Eq:V22}) one easily finds
\begin{equation}
V_{11} = {\cal O}(1),
\qquad
V_{12} = V_{21} = {\cal O}(r),
\qquad
V_{22} = -2\alpha_c^2\frac{\ell+1}{r^2} + {\cal O}(1),
\end{equation}
and from equation~(\ref{Eq:Q}),
\begin{equation}
Q = \frac{\ell + 1}{r} I_2 + {\cal O}(r),
\end{equation}
from which one finally obtains
\begin{equation}
R^T W R 
 = \alpha_c^2\frac{\ell+1}{r^2}\left( \begin{array}{cc} \ell & 0 \\ 0 & \ell+2 \end{array} \right)
 + {\cal O}(1).
\label{Eq:RTWR0}
\end{equation}
Note that the matrix coefficient in front of the $1/r^2$ term on the right-hand side has nonnegative eigenvalues for all $\ell\geq 0$, which is one of the hypothesis in the nodal theorem of reference~\cite{hApQ95}. Furthermore, it follows from the regularity of $\alpha^2/\gamma^2$ at $r = 0$ and from equation~(\ref{Eq:RTWR0}) that the equation ${\cal H}_T v = \sigma^2 v$ has a regular singular point at $r = 0$, with four linearly independent solutions which behave as
\begin{equation}
r^{\ell+1}\left( \begin{array}{c} 1 \\ 0 \end{array} \right),\quad
r^{\ell+2}\left( \begin{array}{c} 0 \\ 1 \end{array} \right),\quad
r^{-\ell}\left( \begin{array}{c} 1 \\ 0 \end{array} \right),\quad
r^{-(\ell+1)}\left( \begin{array}{c} 0 \\ 1 \end{array} \right)
\end{equation}
in the vicinity of $r = 0$. The physical relevant ones (i.e. those leading to perturbations $\delta\varphi_{\ell1}$, $\delta\varphi_{\ell2}$, $\delta\nu$ and $\delta\lambda$ that remain finite near the origin) give rise to the two-parameter family of solutions
\begin{equation}
v \simeq r^{\ell+1}\left( \begin{array}{c} \beta_1 \\ \beta_2 r \end{array} \right),\quad\hbox{or}\quad
f \simeq \left( \begin{array}{c} b_1 \\ b_2 r \end{array} \right),
\label{Eq:LeftBC}
\end{equation}
with free constants $\beta_i$ and $b_i$. In terms of the fields $\delta\varphi_{\ell 1}$ and $\delta\lambda$ appearing in the original system~(\ref{Eq:deltaphi},\ref{Eq:deltalambda}) this leads to local solutions of the form (cf. equation~(\ref{Eq:fDef}))
\begin{subequations}\label{Eq.pulsation.r0}
\begin{eqnarray}
\delta\varphi_{\ell 1} &=& f_1 \simeq b_1,
\label{Eq:r0Expphi}\\
\delta\lambda &=& 2\kappa_\ell r\psi_\ell^2\left[ \omega f_2 + \frac{\psi_\ell'}{\psi_\ell} f_1 \right]
 \simeq 2\kappa_\ell a_0^2 r^{2\ell}[ \ell b_1 + \tilde{b}_2 r^2 ],
\label{Eq:r0ExpLambda}
\end{eqnarray}
\end{subequations}
with free constants $b_1$ and $\tilde{b}_2$. Note that the latter appears only in the $r^2$ correction term in the expansion. A consistent expansion (needed for the numerical implementation in the next section) which includes the $r^2$ and $r^3$ correction terms in both fields $(\delta\varphi_{\ell 1},\delta\lambda)$ will be given in the appendix. Finally, we note that the conditions~(\ref{Eq:LeftBC}) guarantee that the boundary terms in equations~(\ref{Eq.perturbation.number}) and (\ref{Eq:SymId}) vanish at $r = 0$.

\subsection{Asymptotic behavior at $r\to \infty$}\label{Sec:asymptotic}

As $r\to \infty$, the background metric fields $\alpha$ and $\gamma$ both converge to one while the background scalar field quantity $\psi_\ell$ decays exponentially to zero. Therefore, for large $r$ one can replace $\alpha$ and $\gamma$ by their Schwarzschild values,
\begin{equation}
\alpha^2 \simeq \frac{1}{\gamma^2} \simeq 1 - \frac{2M_T}{r},
\end{equation}
with $M_T$ the total mass. (The error in these formulae is exponentially small, as follows from the background equations~(7a) and (7b) in~\cite{Alcubierre:2018ahf}.) Using these expressions in the background equation~(7c) in~\cite{Alcubierre:2018ahf} it then follows that $\psi_\ell$ has the following asymptotic expansion at $r\to \infty$ (note that $\omega^2 < \mu^2$):
\begin{equation}
\psi_\ell(r) = K e^{-\sqrt{\mu^2 - \omega^2} r} r^{-p}\left[ 1 + {\cal O}\left( \frac{1}{r} \right) \right],
\qquad
p = 1 + \frac{M_T(\mu^2 - 2\omega^2)}{\sqrt{\mu^2 - \omega^2}},
\end{equation}
for some constant $K\neq 0$. In particular, this implies that
\begin{equation}
\frac{\psi_\ell'}{\psi_\ell} = -\sqrt{\mu^2 - \omega^2} - \frac{p}{r} + {\cal O}\left( \frac{1}{r^2} \right),
\qquad
\frac{\psi_\ell''}{\psi_\ell} = \mu^2 - \omega^2 + 2\sqrt{\mu^2 - \omega^2}\frac{p}{r}
 + {\cal O}\left( \frac{1}{r^2} \right),
\end{equation}
from which
\begin{equation}
\frac{\alpha^2}{\gamma^2} V_{11} = -4\omega^2 + {\cal O}\left( \frac{1}{r^2} \right),\qquad
V_{12} = V_{21} = V_{22} =  {\cal O}\left( \frac{1}{r^2} \right),
\end{equation}
and
\begin{equation}
\alpha^2\frac{(r\psi_\ell)'}{r\psi_\ell} = -\sqrt{\mu^2 - \omega^2} 
 +\frac{M_T\mu^2}{\sqrt{\mu^2 - \omega^2}}\frac{1}{r} + {\cal O}\left( \frac{1}{r^2} \right).
\end{equation}
Therefore,
\begin{equation}
W =  \left( \begin{array}{cc} \mu^2 + 2\omega^2 & 2\omega\sqrt{\mu^2-\omega^2} \\ 
2\omega\sqrt{\mu^2-\omega^2} & \mu^2 - 2\omega^2 \end{array} \right)
 - \frac{2M_T\mu^2}{r}
 \left( \begin{array}{cc} 1& \frac{\omega}{\sqrt{\mu^2 - \omega^2}} \\ 
 \frac{\omega}{\sqrt{\mu^2 - \omega^2}} & 1 \end{array} \right)
+ {\cal O}\left( \frac{1}{r^2} \right)
\label{Eq:WAsymp}
\end{equation}
in equation~(\ref{Eq:W}), and it follows that the potential term $R^T W R$ of the Schr\"odinger-type operator~(\ref{Eq:HT}) is uniformly bounded for large $r$, which is also a necessary condition to apply the nodal theorem.

Using equation~(\ref{Eq:WAsymp}) and the fact that $r = {\cal O}(r_*)$ one can rewrite the eigenvalue problem $\sigma^2 v = {\cal H}_T v$ in the form $\sigma^2(R v) = R{\cal H}_T R^T (Rv)$ with
\begin{eqnarray}
R{\cal H}_T R^T &=& -\frac{d^2}{dr_*^2} 
 + \left[ 2\omega \left( \begin{array}{rr} 0 & -1 \\ 1 & 0 \end{array} \right)
  + {\cal O}\left( \frac{1}{r_*^2} \right) \right] \frac{d}{dr_*}
\nonumber\\
 &+& \left( \begin{array}{cc} \mu^2 + 3\omega^2 & 2\omega\sqrt{\mu^2-\omega^2} \\ 
2\omega\sqrt{\mu^2-\omega^2} & \mu^2 - \omega^2 \end{array} \right)
 - \frac{2M_T\mu^2}{r_*}
 \left( \begin{array}{cc} 1& \frac{\omega}{\sqrt{\mu^2 - \omega^2}} \\ 
 \frac{\omega}{\sqrt{\mu^2 - \omega^2}} & 1 \end{array} \right)
 + {\cal O}\left( \frac{1}{r_*^2} \right).
\label{Eq:Winf}
\end{eqnarray}
For given values of $\sigma$, this problem has asymptotic solutions of the form
\begin{equation}
R v = e^{\Lambda r_*} r_*^{-P} \left[ {\bf e}_0 + {\cal O}\left( \frac{1}{r_*} \right) \right],
\label{Eq:Rv}
\end{equation}
with a nonvanishing two-vector ${\bf e}_0$, an exponential factor $\Lambda$ and a constant $2\times 2$ matrix $P$. Here, $r_*^{-P}$ refers to the matrix $\exp( -\log(r_*) P)$, and its presence is needed in order to eliminate the $1/r_*$ term appearing in equation~(\ref{Eq:Winf}). Introducing the ansatz~(\ref{Eq:Rv}) into the equation $\sigma^2(R v) = R{\cal H}_T R^T (Rv)$, one finds to leading order:
\begin{equation}
\left[
(\mu^2 + \omega^2 - \sigma^2 - \Lambda^2) I_2 + 2\omega
\left( \begin{array}{cc} \omega & \sqrt{\mu^2-\omega^2} - \Lambda \\ 
\sqrt{\mu^2-\omega^2} + \Lambda & -\omega \end{array} \right) \right] {\bf e}_0 = 0.
\end{equation}
For a nontrivial solution to exist, the determinant of the matrix appearing on the left-hand side must be zero, which yields
\begin{equation}
\Lambda^2 = \mu^2 - \omega^2 - \sigma^2 \pm 2\omega\sigma 
 = \mu^2 - (\omega \mp \sigma)^2,
\label{Eq:Lambda2}
\end{equation}
with corresponding two-vectors proportional to
\begin{equation}
{\bf e}_0 = \left(\begin{array}{cc} \pm\sigma \\ \Lambda + \sqrt{\mu^2 - \omega^2} \end{array}
\right),\qquad
\hbox{or}\quad
{\bf e}_0 =  \left(\begin{array}{cc} \Lambda - \sqrt{\mu^2 - \omega^2} \\ 2\omega\mp \sigma \end{array}
\right).
\label{Eq:e0}
\end{equation}
(These expressions are equivalent to each other as long as $\sigma\neq 0,\pm 2\omega$. When $\sigma = \pm 2\omega$, the first expression should be used, for $\sigma = 0$ the second one.)

Equation~(\ref{Eq:Lambda2}) yields four solutions for $\Lambda$, which, in general, lie in the complex plane. We are particularly interested in understanding the behavior of the real part of these roots, since they determine whether or not the corresponding mode solution decays as $r\to \infty$. To analyze this, we first notice that $\sigma^2$ must be real, since it is an eigenvalue of a self-adjoint operator. Thus, $\sigma$ must either lie on the real or on the imaginary axis of the complex plane. At the intersection, $\sigma = 0$, we have $\Lambda = \pm \sqrt{\mu^2 - \omega^2}$, so in this case the roots are real, with one degenerated positive and one degenerated negative root. By continuity, for small enough values of $|\sigma|$ there are two roots with positive and two roots with negative real parts. More precisely, when $|\mu\sigma| \ll \mu^2 - \omega^2$ one finds
\begin{equation}
\Lambda = \pm\sqrt{\mu^2 - \omega^2} \left[1 \pm \frac{\omega}{\mu^2 - \omega^2}\sigma 
- \frac{\mu^2}{2(\mu^2 - \omega^2)^2}\sigma^2 + {\cal O}(\sigma^3)\right],
\label{Eq:Lambdasm}
\end{equation}
where all four combinations of the signs are possible, the $\pm$ sign choice inside the square parenthesis corresponding to the $\pm$ sign in equation~(\ref{Eq:Lambda2}).

\begin{figure}[t!]
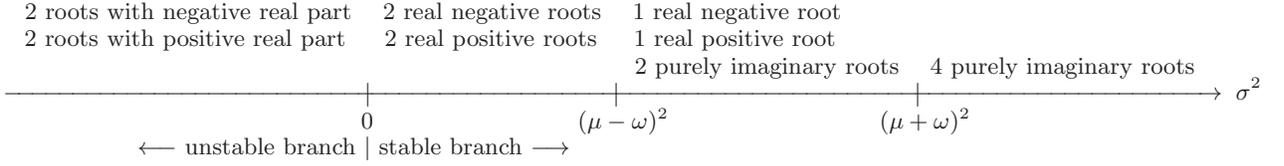

\hspace{-2.3cm} 2 roots with negative real part \hspace{0.2cm} 2 real negative roots \hspace{0.2cm} 1 real negative root \hspace{3cm} 

\hspace{-2.37cm} 2 roots with positive real part \hspace{0.3cm} 2 real positive roots \hspace{0.27cm} 1 real positive root \hspace{3cm} 

\hspace{7.4cm} 2 purely imaginary roots \hspace{0.2cm} 4 purely imaginary roots

\vspace{-.7cm}
\[ \xrightarrow{\hspace*{16cm}} \; \sigma^2 \]

\vspace{-.65cm}
\hspace{0.6cm}$|$ \hspace{3.0cm} $|$ \hspace{3.7cm} $|\quad$\\
\hspace{0.88cm}0 \hspace{2.5cm} $(\mu-\omega)^2$ \hspace{2.59cm} $(\mu+\omega)^2$\\ 
\vspace{-.03cm}
\hspace{-7.52cm} $\longleftarrow$ unstable branch $|$ stable branch $\longrightarrow$
  \caption{Properties of the roots $\Lambda$ of equation~(\ref{Eq:Lambda2}) as a function of the eigenvalue $\sigma^2$ of the operator $\mathcal{H}_T$. Notice that the zero mode $\sigma=0$ divides the stable,  $\sigma^2\ge 0$, from the unstable branch, $\sigma^2< 0$. The number of roots with negative real parts leading to exponential decay of the eigenfunctions $v$ depends on the value of $\sigma^2$. Only for purely imaginary values of $\sigma$ (i.e. $\sigma^2 < 0$) or for real values of $\sigma$ lying in the the interval $-(\mu+\omega)<\sigma<\mu+\omega$ (i.e. $0\leq \sigma^2 < (\mu + \omega)^2$) does one have modes that decay exponentially to zero at spatial infinity.}
\label{Fig:Roots}
\end{figure}

Next, it follows from equation~(\ref{Eq:Lambda2}) and the fact that $\omega^2 < \mu^2$ that the real part of $\Lambda$ cannot vanish if $\sigma$ is purely imaginary. Since $\Lambda$ 
depends continuously on $\sigma$, we conclude there are two roots with positive and two roots with negative real parts as long as $\sigma$ is purely imaginary. On the other hand, we notice that $\Lambda^2$ is real when $\sigma$ is real. Assuming without loss of generality that $\sigma > 0$ and that $0 < \omega < \mu$, it follows from equation~(\ref{Eq:Lambda2}) that 
$\Lambda^2 = 0$ when $\sigma = \mu + \omega$ for the upper sign and $\sigma = \mu - \omega$ for the lower sign. Hence, for $0 < \sigma < \mu - \omega$ there are two roots with 
negative real part and two with positive real part, when $\mu - \omega < \sigma < \mu + \omega$ there is one root with negative and one with positive real part, the remaining two 
roots being purely imaginary, while for $\sigma > \mu + \omega$ it follows that all the roots are purely imaginary, giving rise to oscillatory modes. We summarize these findings in figure~\ref{Fig:Roots}.

Translating these results to the function $f = (f_1,f_2)$ defined in equation~(\ref{Eq:fDef}) by means of the transformation~(\ref{Eq:Transform}), this yields the asymptotic behavior
\begin{equation}
f \sim e^{(\sqrt{\mu^2 -\omega^2} + \Lambda) r} {\bf e}_0.
\end{equation}
For small positive $\sigma^2 > 0$ it follows from equation~(\ref{Eq:Lambdasm}) that there are three exponentially growing modes and one exponentially decaying mode. By analyzing the sign of the term $-\sigma^2 \pm 2\omega\sigma = -\sigma(\sigma\mp 2\omega)$ in equation~(\ref{Eq:Lambda2}) it is not difficult to show that this behavior persists for $0 < \sigma < 2\omega$, while for $\sigma > 2\omega$ all four modes grow exponentially. When $\sigma^2 < 0$ the two roots with the positive real parts clearly give rise to exponentially growing modes. The behavior of the remaining two modes are exponentially damped, as can be seen from the quadratic term in equation~(\ref{Eq:Lambdasm}) and the fact the real part of $\sqrt{\mu^2 - \omega^2} + \Lambda$ cannot vanish for purely imaginary $\sigma$ different from zero. Therefore, for such $\sigma$, there are two exponentially growing and two exponentially decaying modes for $f$. For convenience, we summarize this behavior and the behavior of other fields in table~\ref{Tab:ExpModes}.

For completeness, we also provide the result from the next-to-leading order contribution, which yields the following expression for the matrix $P$ in equation~(\ref{Eq:Rv}):
\begin{equation}
P = \frac{M_T\mu^2}{\Lambda^2 + \omega^2} \left( \begin{array}{cc} 
 \Lambda - \frac{\omega^2}{\sqrt{\mu^2-\omega^2}} 
 & -\omega + \frac{\Lambda\omega}{\sqrt{\mu^2-\omega^2}}\\
  \omega + \frac{\Lambda\omega}{\sqrt{\mu^2-\omega^2}}
 & \Lambda + \frac{\omega^2}{\sqrt{\mu^2-\omega^2}}
\end{array} \right).
\end{equation}
When applied to the two-vector in equation~(\ref{Eq:e0}), this gives
\begin{equation}
P{\bf e}_0 = \frac{M_T\mu^2}{\Lambda^2 + \omega^2} \left( \begin{array}{cc} 
 \Lambda + \frac{\omega^2 \mp \omega\sigma}{\sqrt{\mu^2-\omega^2}} & 0 \\
 0 & \Lambda + \frac{\omega^2 \pm \omega\sigma}{\sqrt{\mu^2-\omega^2}}
\end{array} \right){\bf e}_0,
\end{equation}
and the solutions at infinity are
\begin{equation}
R v = e^{\Lambda r_*}  
r_*^{-\frac{M_T\mu^2}{(\Lambda^2 + \omega)\sqrt{\mu^2 - \omega^2}}
(\Lambda\sqrt{\mu^2 - \omega^2} + \omega^2)} 
\left( \begin{array}{cc} 
r_*^{+\frac{M_T\mu^2\omega\sigma}{(\Lambda^2 + \omega)\sqrt{\mu^2 - \omega^2}}} & 0 \\
0 & r_*^{-\frac{M_T\mu^2\omega\sigma}{(\Lambda^2 + \omega)\sqrt{\mu^2 - \omega^2}}}
\end{array} \right)
 \left[ {\bf e}_0 + {\cal O}\left( \frac{1}{r_*} \right) \right],
\label{Eq:RvAsymptotics}
\end{equation}
with $(\Lambda,{\bf e}_0)$ from equations~(\ref{Eq:Lambda2},\ref{Eq:e0}).

To summarize the findings of this section,  for $\sigma^2 < (\mu-\omega)^2$ there are two linearly independent mode solutions of the pulsation equations which are normalizable at infinity. The corresponding $v$ fields decay exponentially to zero as $r\to \infty$ and guarantee that the boundary terms in equations ~(\ref{Eq.perturbation.number}) and (\ref{Eq:SymId}) vanish at $r=\infty$. In the next section we show (through numerical calculations) that for certain values of $\sigma^2$ an appropriate linear combination of these two modes can be matched to the boundary condition at the origin, see equation~(\ref{Eq:LeftBC}), which yields an eigenfunction of the operator ${\cal H}_T$. For $\sigma^2 > (\mu-\omega)^2$, the number of independent mode solutions which are normalizable and cancel the boundary terms in equations~(\ref{Eq.perturbation.number}) and~(\ref{Eq:SymId}) at infinity is less clear, since in this case there are solutions whose asymptotics is given by equation~(\ref{Eq:RvAsymptotics}) with purely imaginary $\Lambda$. These modes have a power-law behavior of the type $\sim r_*^q$ and in principle, one could determine whether or not they are decaying by analyzing the real part of $q$. However, all the numerical eigenvalues found in the next section satisfy $\sigma^2 < (\mu-\omega)^2$, so that we do not pursue this issue further.

\begin{table}
\begin{tabular}{ll}
 \begin{tabular}{|c||c|c|c|}
 \hline
 \multicolumn{4}{|c|}{$\sigma$ purely imaginary} \\
 \hline
 Roots & $f_1$, $f_2$ & $\delta\varphi_{\ell 1}$, $\delta L$ & $\delta M$\\
 \hline
 \hline
 $++$   & \multicolumn{2}{|c|}{g.e.} & g.e. \\
 $+-$   & \multicolumn{2}{|c|}{g.e.} & g.e. \\
 $-+$   & \multicolumn{2}{|c|}{d.e.} & d.e. \\
 $--$   & \multicolumn{2}{|c|}{d.e.} & d.e. \\
 \hline
 \end{tabular}
 \hspace{0.5cm}
  \begin{tabular}{|c||c|c|c|}
 \hline
 \multicolumn{4}{|c|}{$0 < \sigma < 2\omega$} \\
 \hline
 Roots & $f_1$, $f_2$ & $\delta\varphi_{\ell 1}$, $\delta L$ & $\delta M$\\
 \hline
 \hline
 $++$   & \multicolumn{2}{|c|}{g.e.} & g.e.\\
 $+-$   & \multicolumn{2}{|c|}{g.e.} & d.e.\\
 $-+$   & \multicolumn{2}{|c|}{d.e.} & d.e.\\ 
 $--$   & \multicolumn{2}{|c|}{g.e.} & d.e.\\
 \hline
 \end{tabular}
 \hspace{0.5cm}
 \begin{tabular}{|c||c|c|c|}
 \hline
 \multicolumn{4}{|c|}{$\sigma > 2\omega$} \\
 \hline
 Roots & $f_1$, $f_2\,$ & $\delta\varphi_{\ell 1}$, $\delta L$ & $\delta M$\\
 \hline
 \hline
 $++$   & \multicolumn{2}{|c|}{g.e.} & d.e. \\
 $+-$   & \multicolumn{2}{|c|}{g.e.} & d.e. \\
 $-+$   & \multicolumn{2}{|c|}{g.e.} & d.e. \\ 
 $--$   & \multicolumn{2}{|c|}{g.e.} & d.e. \\
 \hline
 \end{tabular}
\end{tabular}
\caption{Asymptotic behavior at spatial infinity of different variables used along the text: $f_1$ and $f_2$ are the fields introduced in section~\ref{sect.pulsation}, $\delta\varphi_{\ell 1}$ are $\delta L$ are the code variables introduced in the appendix, and $\delta M$ is the linearized mass function, see section~\ref{sect.pulsation}. The signs characterizing the roots of $\Lambda$ make reference to the four possible sign choices in equation~(\ref{Eq:Lambdasm}).
Here {\it g.e.} stands for ``grows exponentially'', whereas {\it d.e.} for ``decreases exponentially'', and we have assumed that $0 < \omega < \mu$ as is the case for the background solution.
\label{Tab:ExpModes}
}
\end{table}


\section{Results}
\label{Sec:Results}

After having derived the pulsation equations and discussed their most important properties, in this section we determine the number of negative energy bound states and the eigenvalues of the Schr\"odinger operator ${\cal H}$ by numerical means. Then we compare our results with those obtained from a fully nonlinear numerical evolution of the EKG system~\cite{Alcubierre:2019qnh}.

\subsection{Number of negative energy bound states}

\begin{figure}
\includegraphics[width=0.99\textwidth]{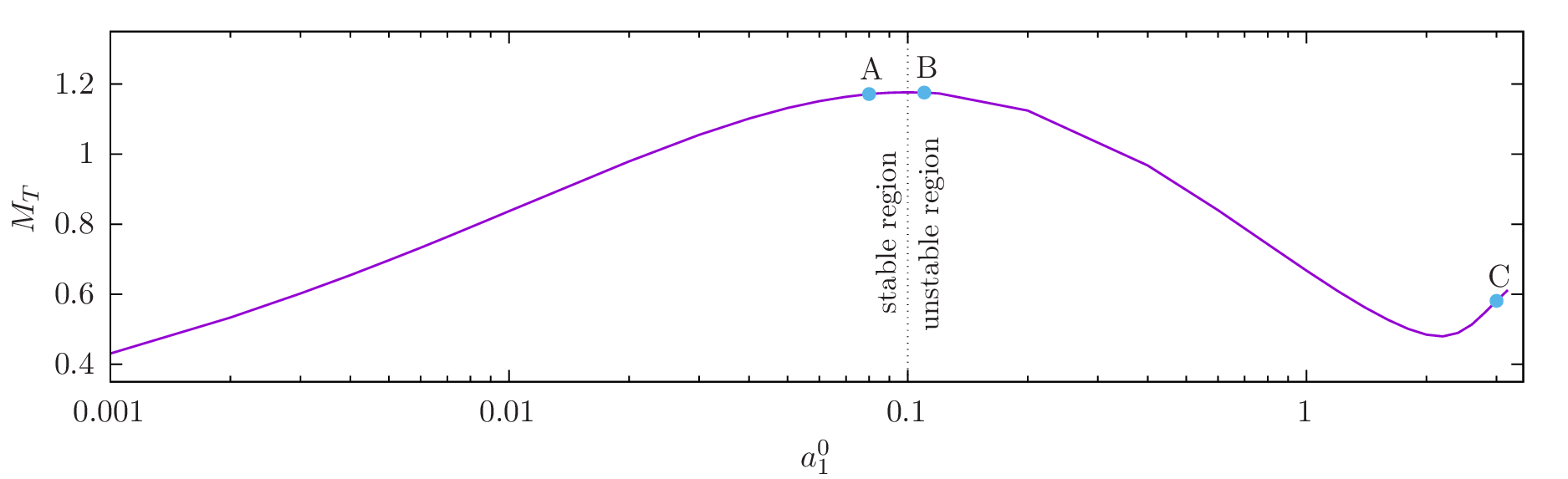}
 \includegraphics[width=0.32\textwidth]{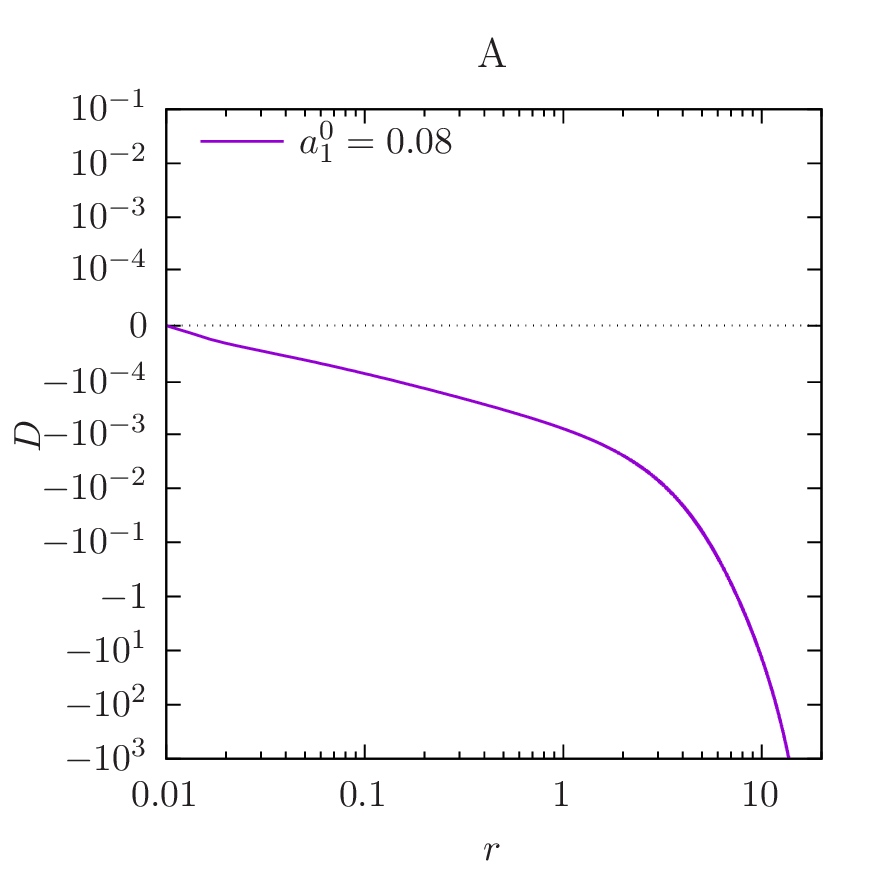}
  \includegraphics[width=0.32\textwidth]{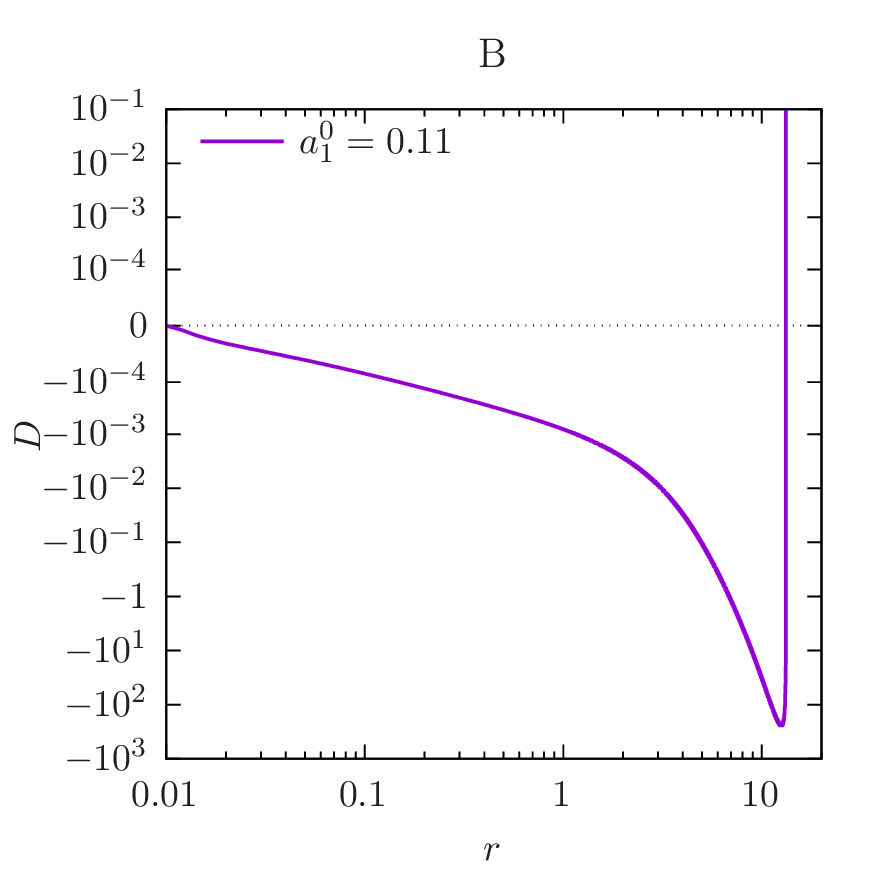}
   \includegraphics[width=0.32\textwidth]{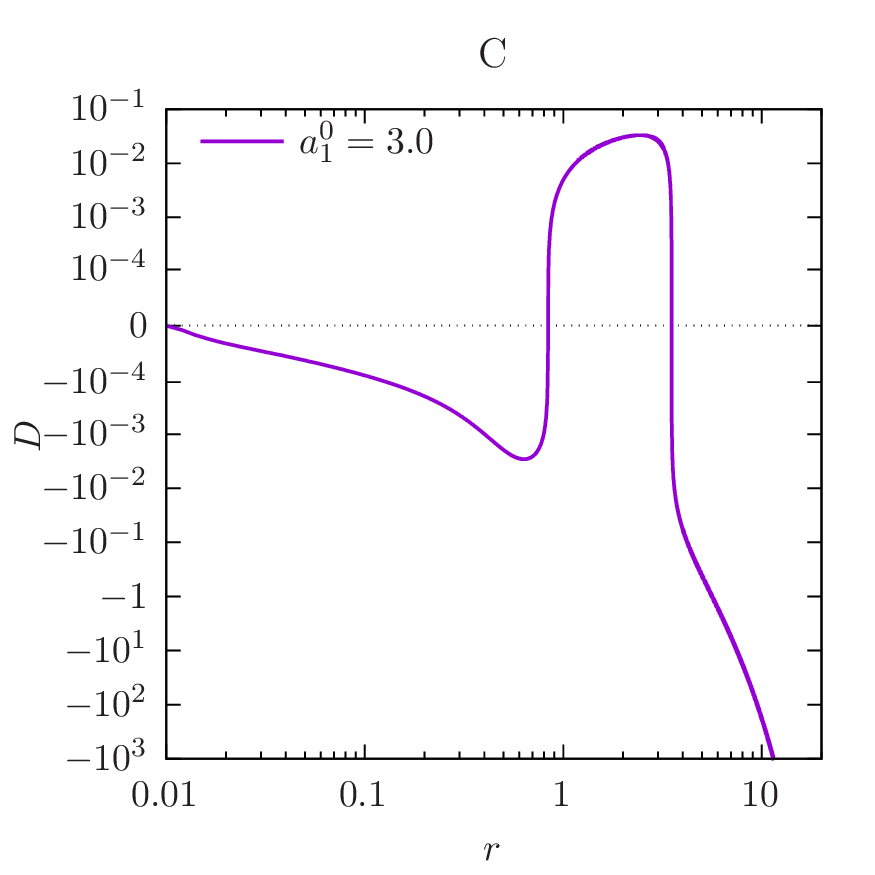}
\caption{The upper panel shows the total mass $M_T$ as a function of the parameter $a_1^0$.  The lower panel shows the determinant function $D(r)$ for the three configurations A ($a_1^0=0.08$), B ($a_1^0=0.11$) and C ($a_1^0=3.0$). Note that the number of zeros of the function $D(r)$ increases by one as one crosses a critical point of the function $M_T(a_{1}^0)$.
}
 \label{fig:teonodal}
\end{figure}

As mentioned in section~\ref{Sec:nodal} the system of equations (\ref{eq:original.system}) forms a self-adjoint coupled system of radial Schr\"odinger equations, once the time-dependence $e^{-i\sigma t}$ is assumed. This fact offers the possibility of studying the point spectrum of the operator $\cal{H}$ given by equation~(\ref{Eq:H}) in a way analogous to a quantum mechanical problem. Indeed, in order to find the number of bound states with negative energy it is sufficient to solve ${\cal H}f=0$ which, as already shown, is equivalent to solving equations~(\ref{eq:original.system}) in the time-independent case $\sigma=0$. The nodal theorem requires finding two solutions $f^{(j)}$ with $j$ labeling the following boundary conditions at $r=r_0$: $f^{(j)}(r_0) = 0$ and $\frac{d}{dr} f^{(j)}(r_0) = {\bf e}_j$, $j=1,2$, with two linearly independent vectors ${\bf e}_1$ and ${\bf e}_2$ in $\Real^2$. Taking these two vectors to be the standard basis in $\Real^2$ and translating the boundary conditions to the fields $\delta\varphi_{\ell 1}$ and $\delta\lambda$ used in our code by means of equations~(\ref{Eq:fDef}), one obtains
\begin{itemize}
    \item For $j=1$ one imposes $\delta\varphi_{\ell1}(r_0)=0$, $\delta\lambda(r_0)=0$, $\delta\varphi_{\ell1}'(r_0)=1$ and $\delta\lambda'(r_0)=2\kappa_\ell r_0 \psi'_\ell(r_0) \psi_\ell(r_0)$.
     \item For $j=2$ one imposes $\delta\varphi_{\ell1}(r_0)=0$, $\delta\lambda(r_0)=0$, $\delta\varphi_{\ell1}'(r_0)=0$ and $\delta\lambda'(r_0)=2\omega \kappa_\ell r_0 \psi_\ell^2(r_0)$.
\end{itemize}

By numerically integrating equations~(\ref{eq:original.system}) with these boundary conditions and $\sigma=0$, one finally computes the determinant
\begin{equation}
    D(r) = \det(f^{(1)}(r),f^{(2)}(r))\,,
\end{equation}
on the interval $(r_0,r_1)$, and by counting the number of zeros of $D(r)$ one finds the number of unstable modes of the corresponding $\ell$-boson star solution. We have tested several choices for $r_0$ and $r_1$ to convince ourselves that the number of zeros of $D(r)$ is independent of the chosen intervals $(r_0,r_1)$.

In figure~\ref{fig:teonodal} we show our results for $\ell$-boson stars with $\ell=1$. The upper panel of this figure shows the relation between the total mass $M_T$ and the parameter $a_\ell^0$ (i.e. the central value of $(2\ell+1)^{-1}r^{-\ell}\psi_\ell(r)$), with the first maximum of the mass at $a_1^0 = a_{1\star}^{0} \approx 0.1$. Recall from the review in section~\ref{Sec:Spherical} that full nonlinear numerical simulations~\cite{Alcubierre:2019qnh} indicate that configurations with $a_1^0 < a_{1\star}^0$ are stable, while those with $a_{1}^0 > a_{1\star}^0$ are unstable. In order to verify that these results are consistent with the nodal theorem, we have chosen one configuration on the stable branch (labeled A and corresponding to $a_1^0 = 0.08$), and two more configurations on the unstable branch (labeled B and C and corresponding to  $a_1^0 = 0.11$ and $a_1^0 = 3.0$, respectively). The behavior of the determinant $D(r)$ is shown for each of these configurations in the lower panel of figure~\ref{fig:teonodal}. Observe that $D(r)$ is a monotonic decreasing function of $r$ for the stable configuration A. No zeros of $D(r)$ are found in this case which proves that this configuration is linearly stable, consistent with the results from the numerical simulations. On the other hand, the function $D(r)$ for configuration B exhibits one zero, showing that it possesses a single unstable mode. Moreover, configuration C, which lies to the right of the local minimum of $M_T$, has two unstable modes as can be seen from the plot of $D(r)$ in this case. The fact that there is an increasing number of unstable modes as $a_\ell^0$ increases and passes through local extrema of the mass will be further discussed in the next subsection, based on the properties of the mode solutions of the pulsation equations.

\subsection{Mode solutions of the pulsation equations}
\label{SubSec:Modes}

We proceed to find the mode solutions (proportional to $e^{-i\sigma t}$) of the pulsation equations~(\ref{eq:original.system}) by means of a numerical shooting algorithm. This system describes an eigenvalue problem for the frequency $\sigma^2$. In order to get a better handle on the solution in the vicinity of $r = 0$, we rescale the variable $\delta\lambda$ by a factor proportional to $r^{-2\ell}$ which is motivated by the asymptotic behavior in equation~(\ref{Eq:r0ExpLambda}). This yields the new system of equations~(\ref{eqs.pulsation.code}) for the fields $\delta\varphi_{\ell 1}$ and $\delta L = \delta \lambda/(2\kappa_\ell\psi_\ell^2)$ derived in the appendix. We numerically integrate this system from $r=0$ outwards, starting with the boundary conditions at $r=0$ given by equations~(\ref{eq.right.boundary.coude}) in the appendix, and fine-tune the values of $\sigma$ and the free parameter $c$ in equation~(\ref{Eq:DvarphiExp0}) until the boundary conditions at $r\to \infty$ discussed in section~\ref{Sec:asymptotic} are satisfied. In practice, this is achieved by choosing an outer radius $r_{\textrm{max}}$, which is the largest radius where the integration is performed, and by imposing the boundary conditions $v_1 = v_2 = 0$ at $r = r_{\textrm{max}}$. In terms of the code variables $\delta\varphi_{\ell 1}$ and $\delta L$ these conditions are equivalent to demanding (since $f = Tv$)
\begin{subequations}\label{Eq.boundary.code}
\begin{eqnarray}
\sqrt{\frac{\gamma}{\alpha}}r_{\textrm{max}}\psi_\ell\delta\varphi_{\ell 1} &=&0\,,
\label{Eq:v1NumBC}\\
\frac{1}{\omega}\sqrt{\frac{\alpha}{\gamma}}\left(\psi_\ell\delta L - r_{\textrm{max}}\psi_\ell'\delta\varphi_{\ell 1}\right)&=&0\,,
\label{Eq:v2NumBC}
\end{eqnarray}
\end{subequations}
at $r_{\textrm{max}}$.\footnote{Theoretically, the boundary conditions~(\ref{Eq.boundary.code}) are equivalent to setting $\delta\varphi_{\ell1} = \delta L = 0$ at $r = r_{\textrm{max}}$. However, we have found that the former work much better in practice.}
In summary, finding the mode solutions of the pulsation equations leads to an eigenvalue problem for $\sigma^2$, for which the solutions must be consistent with both boundary conditions at $r=0$ and $r =r_{\textrm{max}}.$ The double shooting algorithm thus finds both the value of $\sigma^2$ and the value of the free parameter $c$ in equation~(\ref{Eq:DvarphiExp0}) to satisfy correctly the boundary conditions. The value of $r_{\textrm{max}}$ is increased until the values of $\sigma^2$ and $c$ have converged within a tolerance value of $0.1\%$. Some examples showing the values for $\sigma^2$ and $c$ for different configurations with $\ell=1,2$ are shown in table~\ref{Tab:shootingl1}. 

\begin{table*}
\begin{tabular}{|c|c|c|c|c|}
\hline
$\ell$ & $a_\ell^0$ & $\omega$&$\sigma_0^2$ & $c$ \\
\hline\hline
1& $0.050$ & $8.832\times 10^{-1}$ &$3.80\times 10^{-4}$ & $-1.92 \times 10^{-2}$\\
1& $0.080$ & $8.519\times 10^{-1}$ &$2.40 \times 10^{-4}$& $-2.78 \times 10^{-2}$ \\
1& $0.085$ & $8.475\times 10^{-1}$ &$1.91 \times 10^{-4}$& $-2.92 \times 10^{-2}$ \\    
1& $0.090$ & $8.434\times 10^{-1}$ &$1.35 \times 10^{-4}$& $-3.06 \times 10^{-2}$ \\   
1& $0.095$ & $8.394\times 10^{-1}$ &$7.27 \times 10^{-5}$& $-3.20 \times 10^{-2}$\\   
1& $0.100$ & $8.356\times 10^{-1}$ &$3.95 \times 10^{-6}$& $-3.33 \times 10^{-2}$   \\   
1& $0.105$ & $8.320\times 10^{-1}$ &$-7.11 \times 10^{-5}$& $-3.47 \times 10^{-2}$\\   
1& $0.110$ & $8.285\times 10^{-1}$ &$-1.53 \times 10^{-4}$& $-3.60 \times 10^{-2}$\\
\hline
2& $0.006$ & $8.755\times 10^{-4}$ &$2.12\times 10^{-4}$ & $-9.33 \times 10^{-3}$\\
2& $0.008$ & $8.614\times 10^{-1}$  &$1.96 \times 10^{-4}$& $-1.09 \times 10^{-3}$ \\
2& $0.010$ & $8.498\times 10^{-1}$  &$1.61 \times 10^{-4}$& $-1.24 \times 10^{-2}$ \\    
2& $0.012$ & $8.400\times 10^{-1}$ &$1.11 \times 10^{-4}$& $-1.38 \times 10^{-2}$ \\   
2& $0.014$ & $8.316\times 10^{-1}$ &$4.95 \times 10^{-5}$& $-1.51 \times 10^{-2}$\\   
2& $0.016$ & $8.242\times 10^{-1}$ &$-2.21 \times 10^{-5}$& $-1.63 \times 10^{-2}$   \\   
2& $0.018$ & $8.176\times 10^{-1}$ &$-1.02 \times 10^{-4}$& $-1.75 \times 10^{-2}$\\   
2& $0.020$ & $8.117\times 10^{-1}$ &$-1.88 \times 10^{-4}$& $-1.86 \times 10^{-2}$\\
\hline \hline
\end{tabular}
\caption{
The central value $a_\ell^0$, the background frequency $\omega$, the ground state eigenvalue $\sigma^2_0$ and the parameter $c$ in equation~(\ref{Eq:DvarphiExp0}) for some configurations with $\ell=1,2$. Note that at some point the value of $\sigma_0^2$ turns negative signaling the onset of an instability.
}
\label{Tab:shootingl1}
\end{table*}

\begin{figure}
\includegraphics[width=0.49\textwidth]{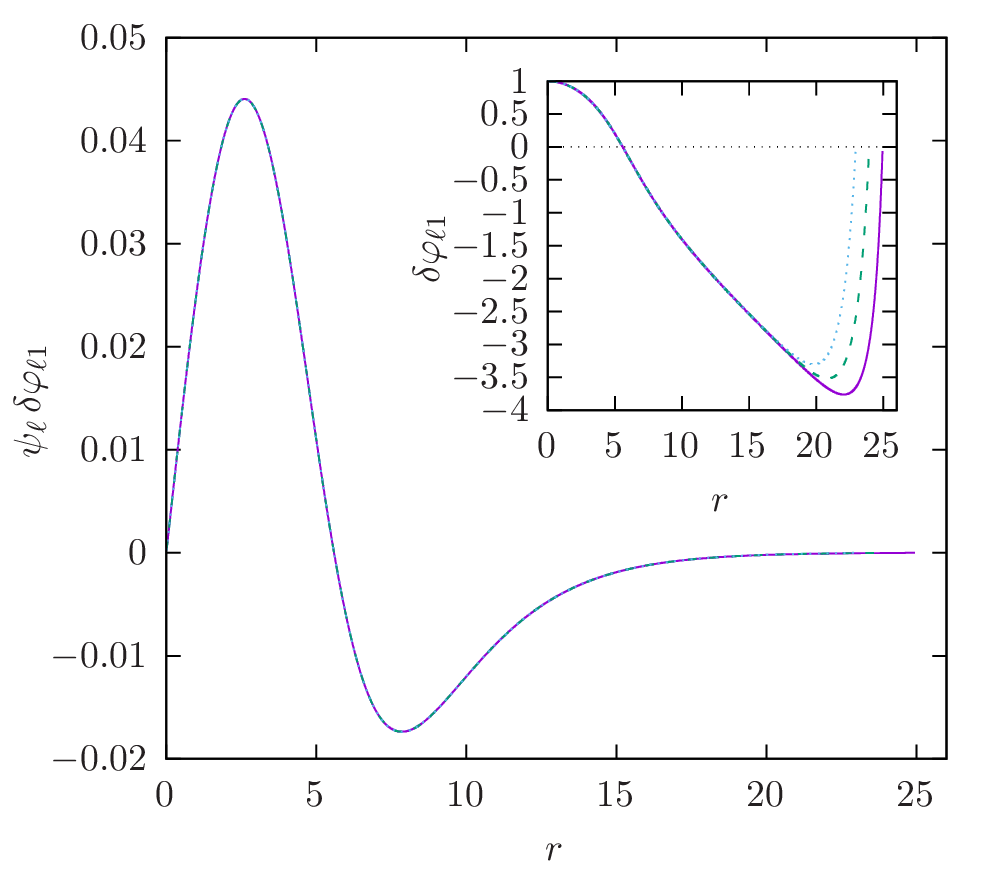}
\includegraphics[width=0.49\textwidth]{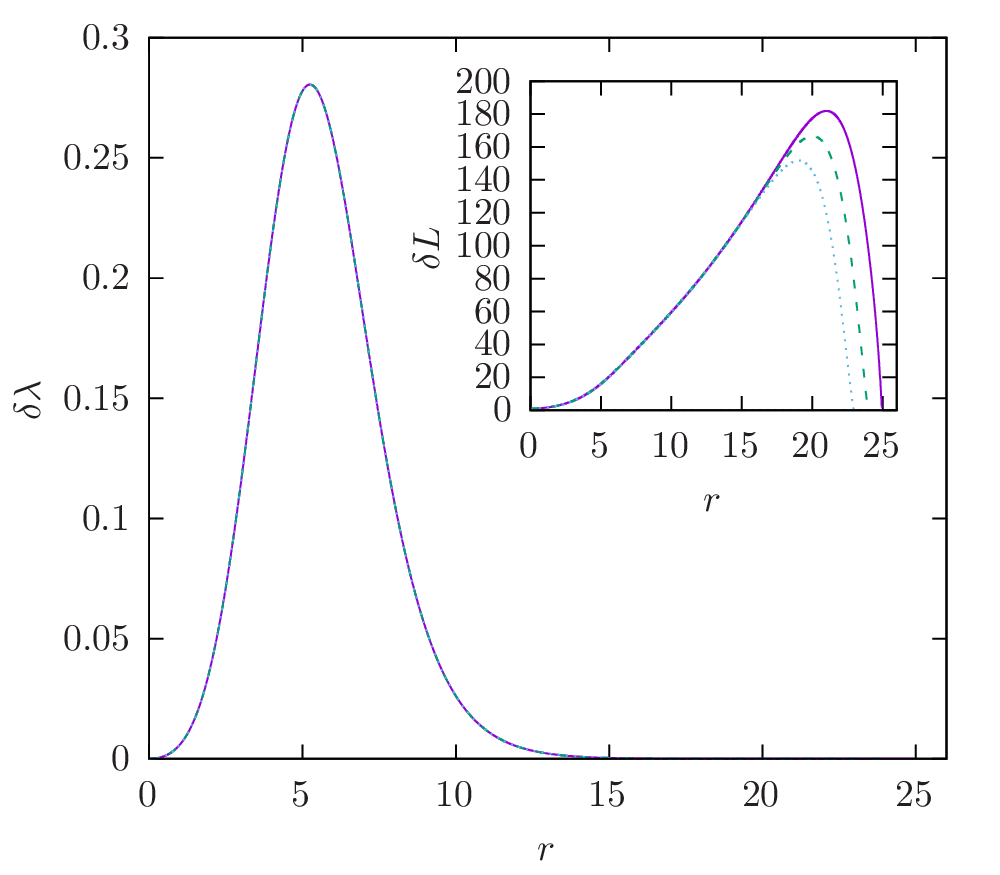}
\caption{
The linearized fields $\psi_\ell\delta\varphi_{\ell1}$ and $\delta\lambda$ as a function of the radial coordinate for the ground state of an $\ell$-boson star with $\ell=1$ and $a_1^0 = 0.08$. Note that the function $\delta\lambda$ has no nodes in the interval $(0,\infty)$. The inner plots correspond to the functions $\delta\varphi_{\ell 1}$ and $\delta L$ from which $\psi_\ell\delta\varphi_{\ell 1}$ and $\delta \lambda$ are computed. The results for different values of $r_{\textrm{max}}$ are shown. In these three cases the required tolerance in the convergence for $\sigma$ has been reached, and consequently the linearized fields $\psi_\ell\delta\varphi_{\ell1}$ and $\delta\lambda$ show no appreciable difference. The differences in the fields $\delta\varphi_{\ell1}$ and $\delta L$ near the boundary are due to the imposition of the boundary conditions~(\ref{Eq.boundary.code}) which, for finite values of $r_{\textrm{max}}$, are equivalent to setting $\delta\varphi_{\ell1} = \delta L = 0$, although in the limit $r_{\textrm{max}}\to \infty$ these quantities are probably exponentially diverging (see table~\ref{Tab:ExpModes}).
}
 \label{Fig:0node}
\end{figure}

\begin{figure}
\includegraphics[width=0.49\textwidth]{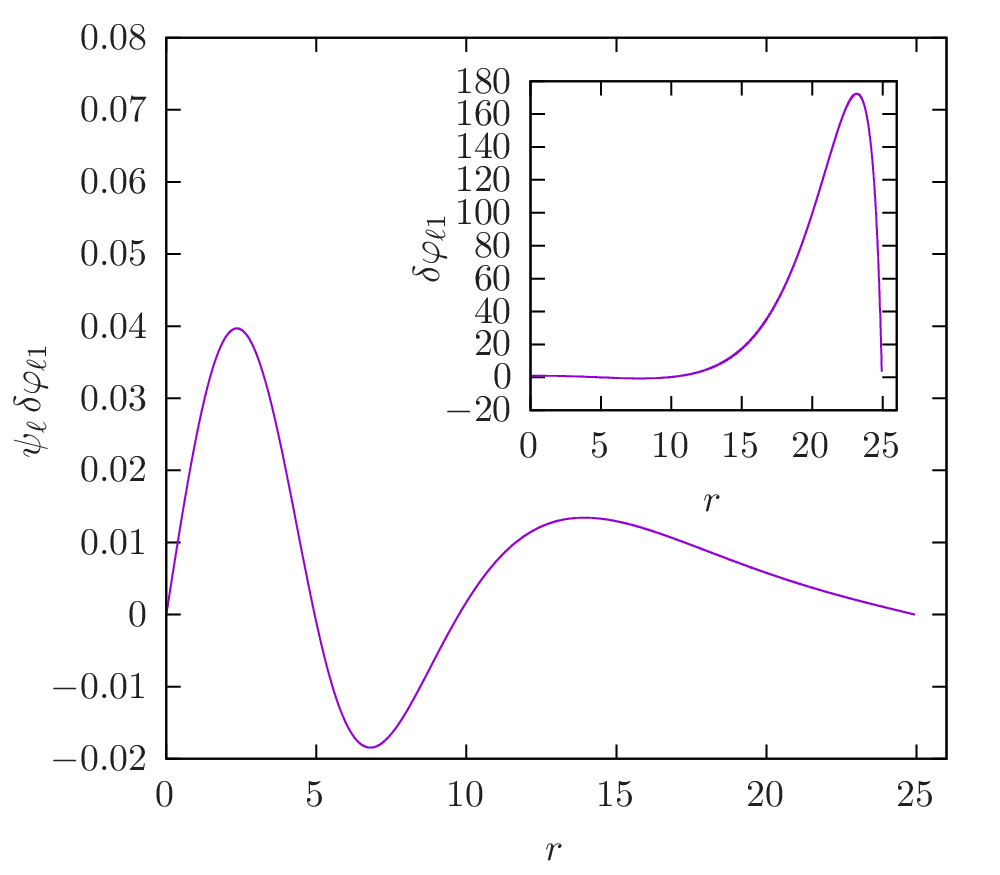}
\includegraphics[width=0.49\textwidth]{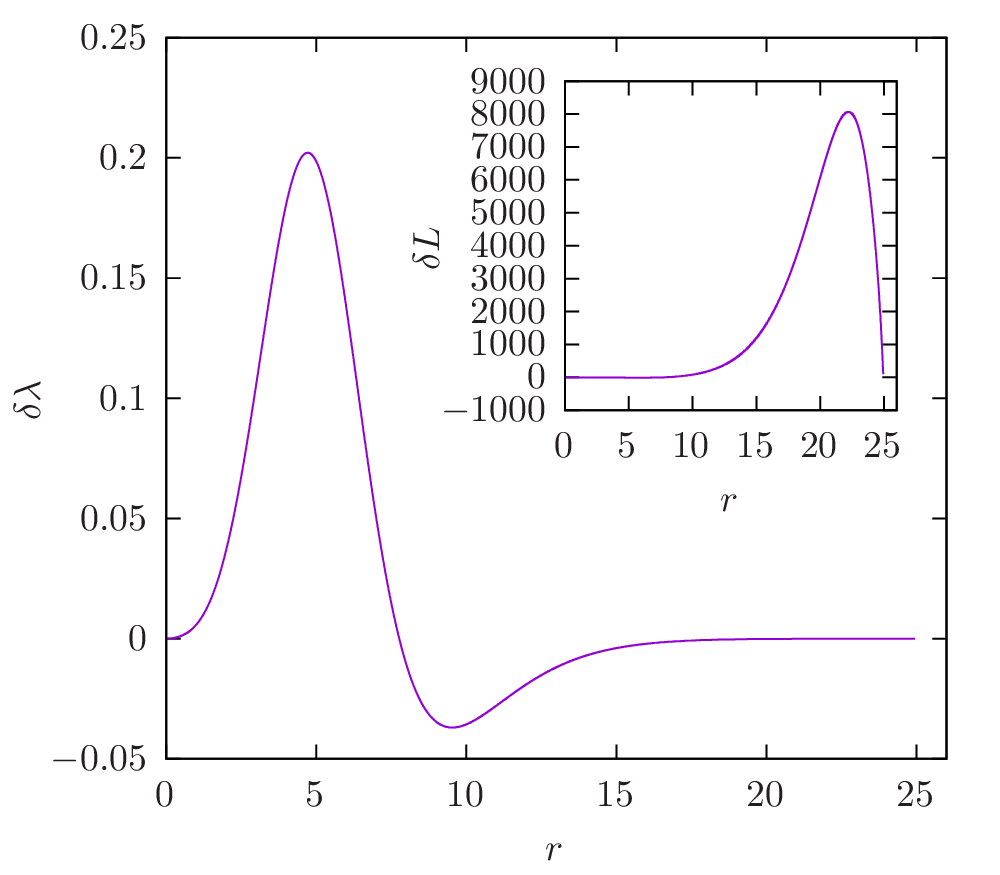}
\caption{
Same as the previous figure but now for the first excited mode with $\sigma_1 = 0.0907$.
}
\label{Fig:1node}
\end{figure}

For each $\ell$-boson star characterized by $\ell$ and $a_\ell^0$ we find a discrete family of mode solutions to the pulsation equations which is characterized by the number of nodes $n$, $n=0,1,2,\ldots$, of the field $\delta\lambda$ in the interval $(0,\infty)$. 
The increasing number of nodes is associated with an increasing value for $\sigma_n^2$, i.e. $\sigma_{n+1}^2 > \sigma_n^2$,
with $\sigma_n$ the frequency of the corresponding mode. Hence, similar to the properties of the background solution describing the $\ell$-boson star, for the solutions of the pulsation equations there are ground, $n=0$, and excited states, $n>0$.
The typical radial dependence of these linear modes is shown in figures~\ref{Fig:0node} and~\ref{Fig:1node} for the particular case of $a_1^0=0.08$ and $\ell=1$. The solution shown in figure~\ref{Fig:0node} is the one with zero nodes in $\delta\lambda$ (the ground state), while the solution with one node in $\delta\lambda$ (the first excited state) is plotted in figure~\ref{Fig:1node}. The radial dependence of the linear modes with $\ell=2,3,..$ and $n=0$ are qualitatively similar to those with $\ell=1$ and $n=0$.

\begin{figure}
\includegraphics[width=0.6\textwidth]{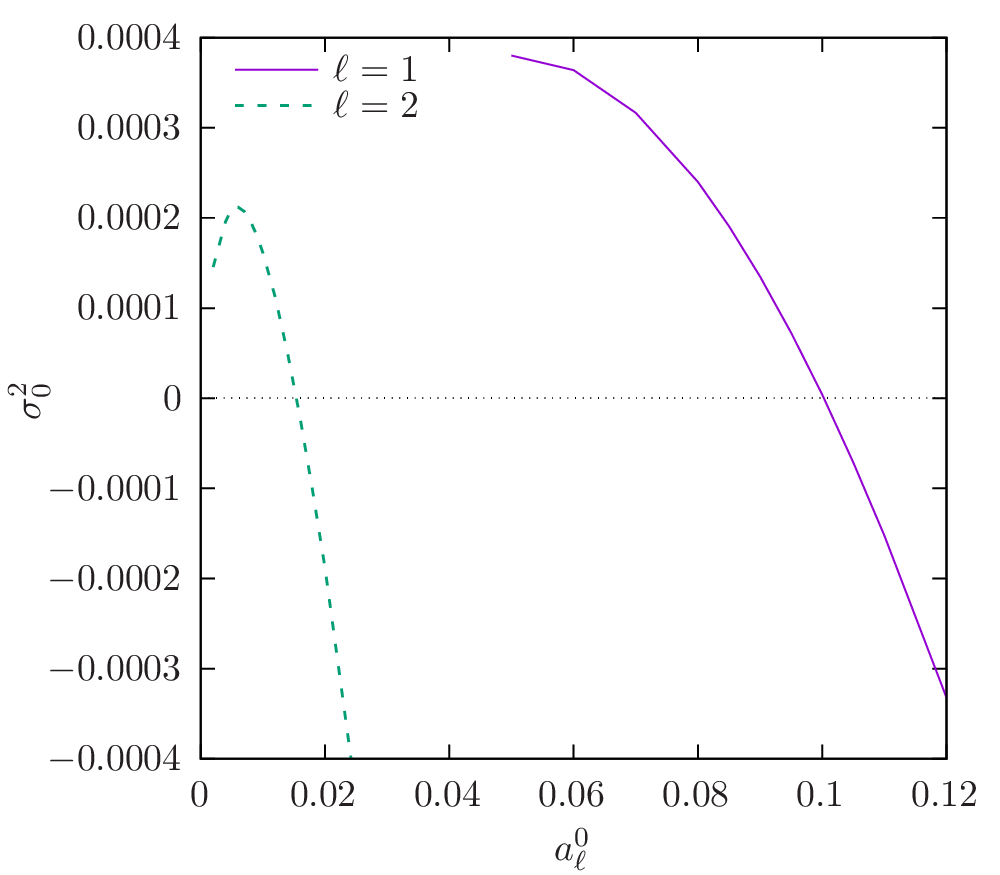}
\caption{The ground state eigenvalues $\sigma_0^2$ as a function of $a_\ell^0$ for $\ell=1$ and $\ell=2$. At some point the eigenvalues turn negative signaling the onset of an instability.
}
 \label{fig:sigma2}
\end{figure}

Next, we compare the eigenvalue $\sigma^2$ for the modes with zero nodes and $\ell=1$ with the corresponding modes with $\ell=2$. The results are shown in figure~\ref{fig:sigma2} where we show the value of $\sigma^2_0$ as a function of $a_\ell^0$.
Table~\ref{Tab:shootingl1} presents the same results as in figure~\ref{fig:sigma2} for a more detailed list of values obtained for $\ell=1,2$. 
As can be seen from figure~\ref{fig:sigma2} and table~\ref{Tab:shootingl1}, the configurations with $a_\ell^0 < a_{\ell\star}^0$ $(\ell=1,2)$ have positive values of $\sigma_0^2 > 0$ which correspond to a pair of oscillating modes which do not grow in time.\footnote{Recall that each eigenvalue $\sigma^2$ gives rise to a pair of mode solutions of the form $\delta\varphi_{\ell 1}(t,r) = e^{\mp i \sigma t} \delta\varphi_{\ell1}(r)$ and $\delta \lambda(t,r)=e^{\mp i\sigma t}\delta \lambda(r)$.} However, when $a_\ell^0 > a_{\ell\star}^0$, $\sigma_0^2 < 0$ is negative which gives rise to a pair of modes, one being exponentially growing in time and the other exponentially damped, showing that the corresponding $\ell$-boson star is linearly unstable. The configuration corresponding to the threshold value $a_\ell^0 = a_{\ell\star}^0$ describes a static mode with $\sigma_0 = 0$, and since by construction our modes are restricted to spherically symmetric configurations which preserve the total mass (and particle number), this mode must represent a linearized solution along the $\ell$-boson star configurations at a point where the first variation of the mass is zero (cf.~\cite{Gleiser:1989a}). In fact, it turns out that the threshold value $a_{\ell\star}^0$ corresponds precisely to the value of the maximum mass configuration. These results explain the reason behind the transition stable-unstable at the first maximum of the mass that we pointed out in figure~\ref{fig:MvsR}, and are fully compatible with the results from the full nonlinear numerical evolution of the $\ell$-boson star configurations performed in~\cite{Alcubierre:2019qnh}.

\begin{figure}
\includegraphics[width=0.79\textwidth]{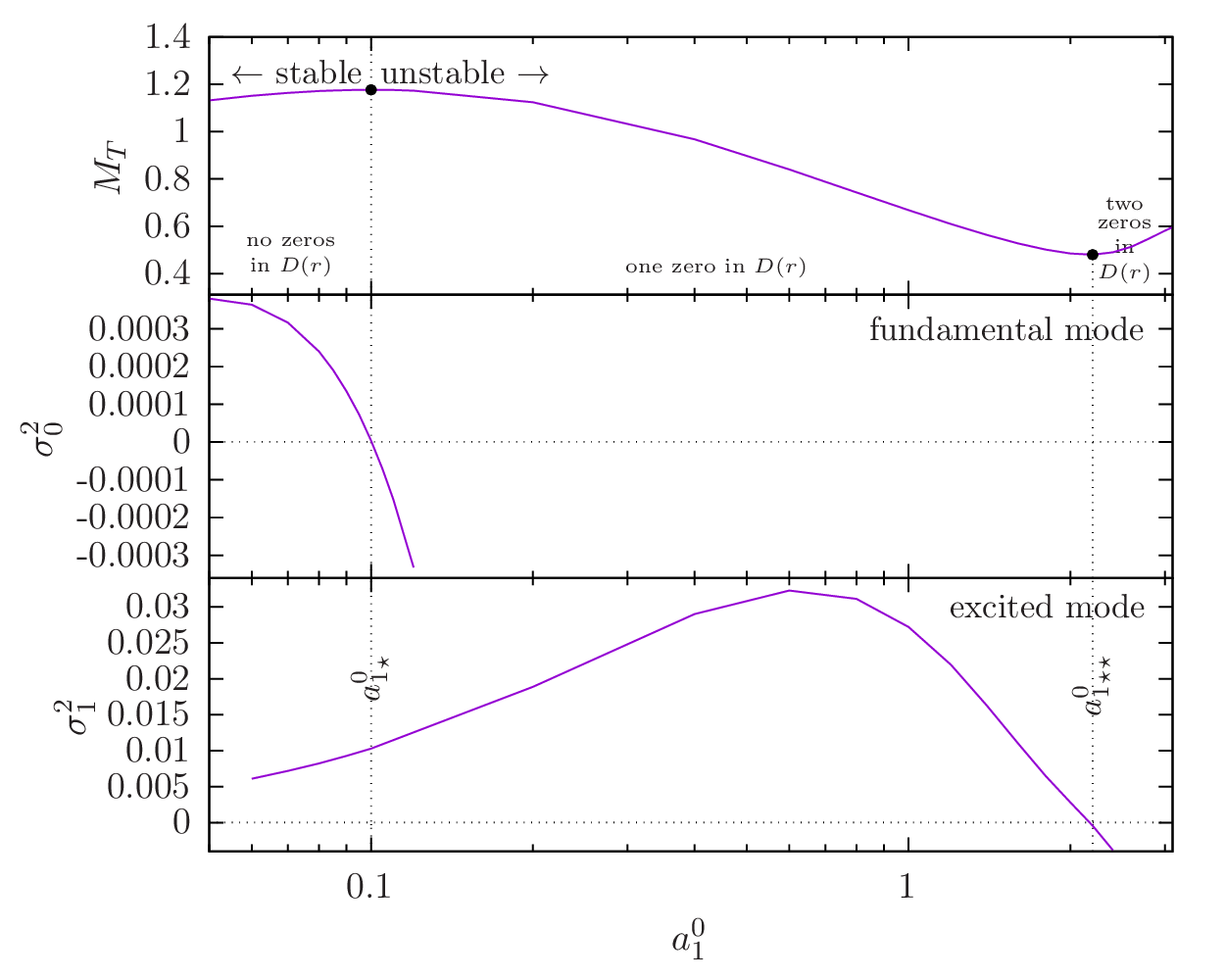}
\caption{The ground state eigenvalue $\sigma^2_0$ and its first excitation $\sigma_1^2$ for $\ell=1$ as a function of $a_\ell^0$. Note that $\sigma_0^2$ remains always lower than $\sigma_1^2$, and that whereas the ground state eigenvalue crosses zero at $a_{1}^0=a_{1\star}^0$, the first excited eigenvalue does so at $a_{1}^0=a_{1\star\star}^0$.}
 \label{fig:zeros}
\end{figure}

We conclude this subsection by commenting on the origin of the number of instabilities that the nodal theorem is counting, based on the mode solutions to the pulsation equations. As we have mentioned, it is possible to find solutions to the pulsation equations with $n=0,1,2,\ldots$ nodes of the function $\delta \lambda$. For instance, in figure~\ref{fig:zeros} we have plotted $\sigma^2_0$ and $\sigma^2_1$ for configurations with $\ell=1$ as a function of $a_1^0$. For comparison, the top panel of figure~\ref{fig:zeros} shows once more the total mass $M_T$ as a function of $a_1^0$. As discussed above, the ground state eigenvalue $\sigma_0^2$ is positive for $a_1^0 < a_{1\star}^0$, which means that the corresponding $1$-boson star is stable under linear radial perturbation. In fact, the nodal theorem shows that the function $D(r)$ has no zeros for $a_1^0 < a_{\star1}^0$ (cf. configuration A in figure~\ref{fig:teonodal}) showing that there are, in fact, no unstable modes in this region. For $a_1^0 > a_{1\star}^0$, $\sigma_0^2$ becomes negative, the corresponding $1$-boson star is unstable, and the nodal theorem shows that $D(r)$ has at least one zero in this region (cf. configurations B and C in figure~\ref{fig:teonodal}). Next, let us examine the behavior of the first excited eigenvalue $\sigma_1^2$ as a function of $a_1^0$ (see the lower panel of figure~\ref{fig:zeros}). Note that $\sigma_1^2 > 0$ is positive for values $a_1^0  < a_{1\star\star}^0$ where $a_{1\star\star}^0$ corresponds to the configuration for which $M_T$ has its first local minimum. This means that there are no additional unstable modes in this region, a fact that is confirmed by the results from the nodal theorem that show that $D(r)$ has only one zero for $a_{1\star}^0 < a_1^0  < {a_{1\star\star}^0}$ (cf. configurations B in figure~\ref{fig:teonodal}). However, for $a_1^0 > a_{1\star\star}^0$ the eigenvalue $\sigma_1^2$ becomes negative meaning that in this region the configuration has \emph{two} exponentially in time growing modes (one which is due to $\sigma_0^2 < 0$ and another one due to $\sigma_1^2 < 0$), a fact that is again confirmed by the nodal theorem (cf. configurations C in figure~\ref{fig:teonodal}). From this picture, we conjecture that as $a_1^0$ increases, the corresponding $1$-boson star  acquires one additional unstable mode each time $a_1^0$ crosses an extremum of the mass.

\subsection{Comparison of linear perturbation theory with numerical perturbations of $\ell$-boson stars}

In this subsection we compare the results of the perturbation theory with those of a full non-linear numerical evolution of a perturbed $\ell$-boson star in the stable branch.

For our perturbed initial data we follow the procedure described in~\cite{Alcubierre:2019qnh}. We start from a solution corresponding to a stationary $\ell$-boson star and add a small, but finite, perturbation. The perturbation is constructed in such a way that the initial momentum density remains zero, in order to guarantee that the momentum constraint is trivially satisfied.  Also, we choose a perturbation such that the {\em local}\/ boson density does not change at $t=0$, that is we keep  $\im[ \phi_\ell^*\Pi_\ell ]$ constant in equation~\eqref{Eq:ParticleNumber}.  We then solve again the Hamiltonian constraint in order to have fully consistent initial data and evolve the full EKG system numerically in time.  Notice that keeping fixed the local boson density does not in fact keep the total boson number $N_B$ conserved, as the volume element will change once we solve again the Hamiltonian constraint. Basically, when constructing our initial data we are ignoring the term $\delta \lambda$ in equation~\eqref{Eq:deltaN}. This changes slightly the background solution that we need to compare with the results of our numerical evolution, but the effects are so small that they are negligible for practical purposes, as we have corroborated.

We perform a very long time evolution in order to have a large number of oscillations of the system. To find the frequencies of the perturbed $\ell$-boson star, we Fourier transform the value of the lapse function $\alpha_c := \alpha(r=0)$ at the origin. Notice that for an unperturbed star this value should remain constant. Using a fast Fourier transform (FFT), we then compute the power spectrum of $\alpha_c$.

\begin{figure}
\includegraphics[width=0.49\textwidth]{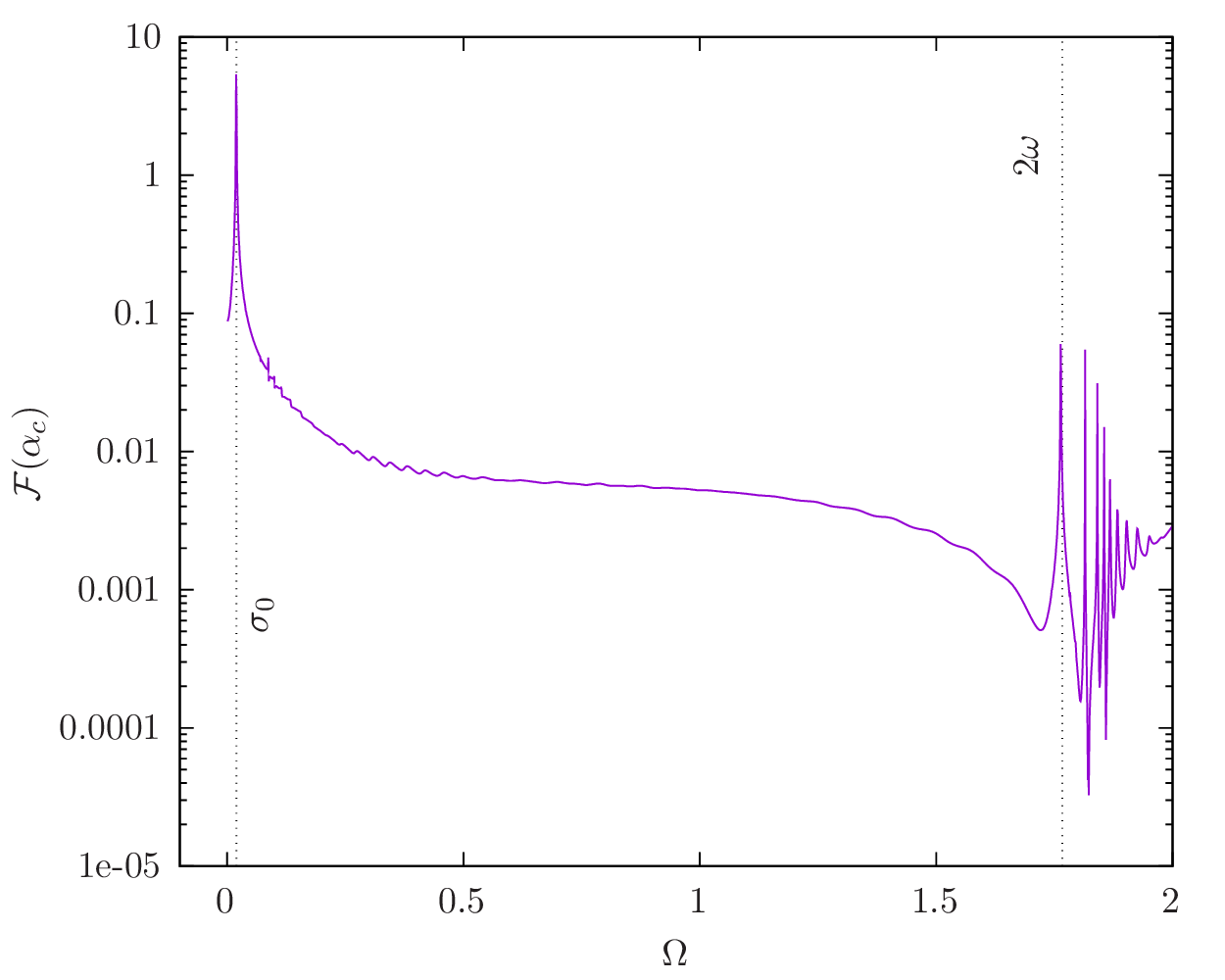}
\includegraphics[width=0.49\textwidth]{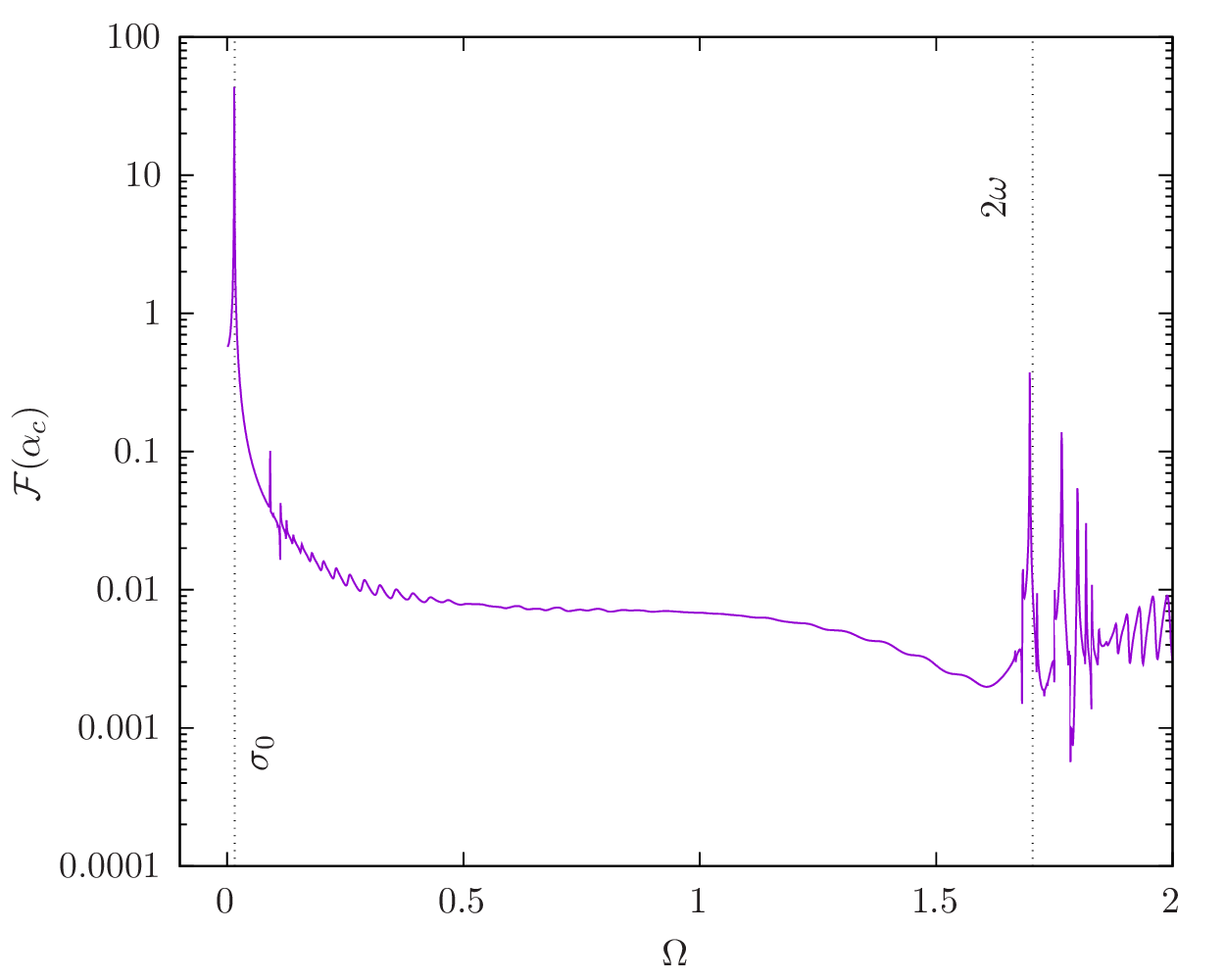}
\caption{Power spectrum of the evolution of the value of the lapse function at the origin for perturbed $\ell$-boson stars with $a^0_1=0.05$ (left) and $a^0_1=0.08$ (right), both with $\ell=1$.  For details of the evolutions see \cite{Alcubierre:2019qnh}. The vertical lines on the left of both panels correspond to the values of $\sigma_0$ found in this work by means of the linear perturbation analysis (see table~\ref{Tab:shootingl1}). The vertical lines on the right correspond to twice the frequency of the fundamental mode $\omega_0$. Note that the dominant peak matches very well the predicted frequency $\sigma_0$ from the linear perturbation theory. Other peaks seem to correspond to combinations of excited modes $\omega_j$ and half the value of $\sigma_0$, although we have no current explanation for this.
}
 \label{fig:fourier}
\end{figure}

Figure~\ref{fig:fourier} shows the power spectrum of $\alpha_c$ for the time interval $[0,20 000]$ for the case of a perturbed $\ell$-boson star with $\ell=1$, and two different values of $a^0_1$ corresponding to the background configurations in the first two rows of table~\ref{Tab:shootingl1}. 
The power spectrum clearly indicates there are several frequencies at which the system oscillates. The largest (leftmost) peak in both cases corresponds to the value of $\sigma_0$ whose square is given in table~\ref{Tab:shootingl1},
that is the fundamental mode of the perturbation analysis. The second dominant frequency corresponds to twice the frequency of the original unperturbed $\ell$-boson star. The reason for this is that the lapse function (as well as other metric quantities) depends on the energy density of the field, which is proportional to its square modulus, so it oscillates with twice the frequency.

The other peaks present in the power spectrum seem to correspond to linear combinations of twice the values of the overtones of the $\ell$-boson star (or the first two excited states) and halves of  $\sigma_0$. At this point we do not have a clear understanding of exactly why these particular combinations appear in the spectrum, but they would seem to originate in the nonlinear coupling among the different modes and depend on the initial perturbation. A further analysis of the exact modes that are excited is beyond the scope of the present work.

\section{Conclusions}
\label{Sec:Conclusions}

$\ell$-boson stars~\cite{Alcubierre:2018ahf} are a generalization of the standard boson stars with $\ell=0$ which are obtained from a collection of an arbitrary odd number $N=2 \ell+1$ of complex massive scalar fields with an internal $U(N)$ symmetry. Even if spherical as a whole, the individual scalar fields are described by an eigenstate of the angular momentum operator with angular momentum number $\ell$, with fields belonging to different azimuthal number $m$ having the same amplitude $\psi_\ell(r)$. The resulting configurations are described by regular, asymptotically flat, static and spherically symmetric solutions to the classical EKG system, and they can be parametrized by $\ell$, an excitation number $n=0,1,2,\ldots$ that counts the number of nodes of the wave function, and a continuous finite parameter $a_\ell^0$ representing the amplitude of the radial function $(2\ell+1)^{-1} r^{-\ell}\psi_\ell(r)$ at the origin.

In order to have a possible phenomenological impact, $\ell$-boson stars need to be dynamically stable as solutions of the EKG equations. In this paper, by means of a linear perturbation analysis, we have studied the stability of $\ell$-boson stars by analyzing the time evolution of linearized radial perturbations which conserve the total number of particles and mass. Our results support the previous findings of our numerical study in~\cite{Alcubierre:2019qnh} by confirming that for a given $\ell$, nodeless $n=0$ $\ell$-boson stars are stable if $a_\ell^0$ is smaller than a critical value $a_{\ell\star}^0$ which corresponds to the maximum mass configuration.

The existence of a branch of solutions that is stable under linear perturbations was confirmed in this work 
by two different methods:
\begin{enumerate}
\item On the one hand we have derived the pulsation equations that describe the evolution of the linearized radial perturbations to the $\ell$-boson star ground state configurations. These equations form a self-adjoint coupled system of Schr\"odinger equations, which is then suitable to the applicability of the generalized nodal theorem~\cite{hApQ95}. This theorem allows one to count the number of unstable modes of the pulsation equations by counting the number of zeros of a certain determinant function $D(r)$ constructed from two independent zero modes. As we have shown, the configurations with $a_\ell^0 < a_{\ell\star}^0$ have no unstable modes while those with $a_\ell^0 > a_{\ell\star}^0$ have one, two or more unstable modes for the cases $\ell=1,2$ we have computed. Furthermore, the number of unstable modes associated with the configuration $a_\ell^0$ coincides with the number of critical points of the function $M_T(a_\ell^0)$ in the interval (0,\ $a_\ell^0$). See figure~\ref{fig:teonodal} for details.
    
\item On the other hand, we have solved the pulsation equations for solutions with a harmonic time dependency of the form $e^{-i\sigma t}$ by means of a numerical shooting algorithm. The mode solutions reveal that the linear radial perturbation of those $\ell$-boson stars with $a_\ell^0 < a_{\ell\star}^0$ oscillates with a real frequency $\sigma$ around the background configuration. In contrast, configurations with $a_\ell^0 > a_{\ell\star}^0$ possess linear modes growing exponential in time, for the cases $\ell=0,1,2$ we have analyzed. In particular, our analysis allows us to compute numerically the frequency $\sigma$ for the stable configuration (see figure~\ref{fig:zeros}), and we have found that this frequency correctly predicts the position of the highest peak in the Fourier spectrum of a  perturbation obtained by numerically evolving the nonlinear EKG system (see figure~\ref{fig:fourier}).
\end{enumerate}

Our results based on the nodal theorem and the computation of the mode solutions of the pulsation equations complement each other. The counting of the number of instabilities makes sure that no unstable modes have been missed in our shooting algorithm, and in particular shows the existence of a stable configurations, whereas the computation of the mode solutions allows one to compute the spectrum of frequencies $\sigma$ associated with the perturbations. Furthermore, the number of instabilities given by the nodal theorem corresponds to the number of the linear perturbations that grow in time exponentially.

Our results confirm that the pulsation equations capture the dynamics of the evolution of $\ell$-boson stars. All the results presented here, together with the numerical studies  presented in \cite{Alcubierre:2019qnh}, suggest that $\ell$-boson stars do have a branch which is stable, at least with respect to radial perturbations, thus making them suitable to model astrophysical objects.

If ultimately proven to be stable also with respect to non-spherical perturbations, confirming the findings in~\cite{Jaramillo:2020rsv}, one would have compact objects made of scalar fields that are stable and that can have compactness larger than  standard $\ell=0$ boson stars, as shown in figure~\ref{fig:MvsR}. The astrophysical implications of these compact structures are still unknown. However, they should be of considerable interest for those dark matter models which are described by a (zero spin) scalar field.


\acknowledgments

This work was partially supported by 
CONACyT Ciencia de Frontera Projects 
No. 376127 ``Sombras, lentes y ondas gravitatorias generadas por objetos
compactos astrof\'\i sicos", and No. 304001 ''Estudio de campos escalares con aplicaciones en cosmolog\'ia y astrof\'isica", as well as
DGAPA-UNAM grants IN110218 and IN105920.
OS was partially supported by a CIC grant to Universidad Michoacana de San Nicol\'as de Hidalgo.
ADT was partially suported by CONACyT grant No. 286897.


\appendix
\section*{Appendix. The pulsation equations in a form more suitable for numerical integration}

As discussed in section~\ref{Sec:Results} the computation of the eigenvalues $\sigma^2$ and the associated linearized modes is based on a numerical shooting algorithm which integrates the pulsation equations outwards starting from the origin $r=0$. To this purpose, and taking into account the asymptotic behavior of the fields near $r=0$, see equation~(\ref{Eq.pulsation.r0}), it is convenient to replace $\delta\lambda$ with the rescaled quantity
\begin{equation}
\delta L := \frac{\delta\lambda}{2\kappa_\ell\psi_\ell^2}.
\label{Eq:deltaL}
\end{equation}
After this rescaling, equations~(\ref{Eq:deltaphi},\ref{Eq:deltalambda}) assume the following form:
\begin{subequations}\label{eqs.pulsation.code}
\begin{eqnarray}
\delta\varphi_{\ell 1}'' &=& -\left( \frac{2}{r} + \frac{\alpha'}{\alpha} - \frac{\gamma'}{\gamma} \right)\delta\varphi_{\ell 1}' - \frac{2}{r} \delta L'
\nonumber\\
 &+& 2\gamma^2\left[ 
 \frac{1}{\gamma^2}\left(  \frac{ u_\ell'}{ u_\ell} + \frac{\ell}{r} \right)^2 
 + \kappa_\ell r^{2\ell}\mu_\ell^2  u_\ell\left( r u_\ell' + \ell u_\ell \right)
  + \mu_\ell^2
  + \frac{2\omega^2 - \sigma^2}{2\alpha^2} \right] \delta\varphi_{\ell 1}
\nonumber\\
  &-& 2\left[ \frac{1}{r^2} + \frac{2}{r}\left(\frac{u_\ell'}{u_\ell} + \frac{\ell}{r} - \frac{\gamma'}{\gamma}  \right) 
  + \kappa_\ell r^{2\ell-1} u_\ell\left( r u_\ell' + \ell u_\ell \right) 
\left( \frac{\alpha'}{\alpha} - \frac{\gamma'}{\gamma} + \frac{u_\ell'}{u_\ell} + \frac{\ell + 1}{r} \right)
- \kappa_\ell\gamma^2 r^{2\ell}u_\ell^2 \left( \mu_\ell^2 - \frac{\omega^2}{\alpha^2} \right) \right]
\delta L,
\nonumber\\
\label{Eq:dphi1}
\end{eqnarray}
\begin{eqnarray}
\delta L'' &=& -2\left[ 2\left( \frac{u_\ell'}{u_\ell} + \frac{\ell}{r} \right) 
 - r\gamma^2 \mu_\ell^2 \right] \delta\varphi_{\ell1}'
 - \left[  4\frac{u_\ell'}{u_\ell} + \frac{4\ell}{r} + 3\left( \frac{\alpha'}{\alpha} - \frac{\gamma'}{\gamma} \right) \right] \delta L'
 \nonumber\\
  &-& 2\gamma^2\left[ \frac{2}{\gamma^2} \left( \frac{u_\ell'}{u_\ell} + \frac{\ell}{r} \right)^2
   - r\mu_\ell^2\left( 2\frac{u'_{\ell}}{u_\ell} +\frac{2\ell}{r} + 2\frac{\alpha'}{\alpha} + \frac{\gamma'}{\gamma} \right) + \frac{\ell(\ell+1)}{r^2} \right] \delta\varphi_{\ell1}
 \nonumber\\
  &+& 2\left[  2\kappa_{\ell}r^{2\ell-2}(r u_\ell' + \ell u_\ell)^2
   - \left( \frac{u_\ell'}{u_\ell}  + \frac{\ell - 1}{r} + \frac{\alpha'}{\alpha} - \frac{\gamma'}{\gamma} \right)^2
   + \frac{2}{r^2} - \left( \frac{4\alpha'}{r\alpha} - \frac{\gamma'}{r\gamma} \right)
   + \left( \frac{\gamma'}{\gamma} \right)'
   - \gamma^2\left( \mu_\ell^2 - \frac{2\omega^2 - \sigma^2}{2\alpha^2} \right)
  \right] \delta L,
\nonumber\\
\label{Eq:dL}
\end{eqnarray}
\end{subequations}
where we have introduced the shortcut notation $\mu_\ell^2 := \mu^2 + \ell(\ell+1)/r^2$ and set $\psi_\ell = r^\ell u_\ell$, taking into account that the background field $\psi_\ell$ scales like $r^\ell$ near $r = 0$, see equation~(\ref{Eq:psi0}). In these expressions, we eliminate the derivatives of the metric coefficients using the background equations~(17a) and (17b) in reference~\cite{Alcubierre:2018ahf} and
\begin{equation}
\left( \frac{\gamma'}{\gamma} \right)' 
 = \frac{\gamma^2-1}{\gamma r}\left( \frac{\gamma}{r} - \gamma' \right)
 + \kappa_\ell r\left\{ \psi_\ell'\psi_\ell'' 
 - \left[ \frac{\omega^2\alpha'}{\alpha^3} + \frac{\ell(\ell+1)}{r^3} \right]\gamma^2\psi_\ell^2
 + \left( \mu_\ell^2 + \frac{\omega^2}{\alpha^2} \right)
 \left(\gamma\gamma'\psi_\ell^2 + \gamma^2\psi_\ell\psi_\ell' \right) \right\},
\end{equation}
where here $\psi_\ell$ and its first two derivatives can be computed from $\psi_\ell = r^\ell u_\ell$ and equation~(17c) in~\cite{Alcubierre:2018ahf}.

The system~(\ref{eqs.pulsation.code}) is singular at $r=0$; however, as we have shown in section~\ref{SubSec:Asym0} there is a two-parameter family of solutions which are regular at the origin. To find the corresponding expansions of these solutions which can be used to start the numerical integration, we write the system~(\ref{eqs.pulsation.code}) in the form
\begin{equation}
V'' = F(r) V' + G(r) V,\qquad
V := \left( \begin{array}{c} \delta\varphi_{\ell 1} \\ \delta L \end{array} \right),
\label{Eq:V}
\end{equation}
and look for solutions of the form
\begin{equation}
V(r) = V_0 + V_2 r^2 + V_4 r^4 + {\cal O}(r^6),
\label{Eq:VExpansion}
\end{equation}
near $r = 0$. It follows from equations~(\ref{Eqs.behavior.origin}) that the matrix-valued functions $F(r)$ and $G(r)$ have expansions of the following form:
\begin{equation}
F(r) = \frac{F_{-1}}{r} + F_1 r + {\cal O}(r^3),\qquad
G(r) = \frac{G_{-2}}{r^2} + G_0 + G_2 r^2 + {\cal O}(r^4),
\end{equation}
with constant matrices $F_i$ and $G_i$. Substituting the expansion~(\ref{Eq:VExpansion}) into equation~(\ref{Eq:V}) leads to
\begin{equation}
G_{-2} V_0 = 0,\qquad
(2I - 2F_{-1} - G_{-2}) V_2 = G_0 V_0,\qquad
(12 I - 4F_{-1} - G_{-2}) V_4 = (2F_1 + G_0)V_2 + G_2 V_0.
\label{Eq:MatrixRels}
\end{equation}
Since
\begin{equation}
F_{-1} = -2\left( \begin{array}{cc} 1 & 1 \\ -\ell(\ell-1) & 2\ell \end{array} \right),\qquad
G_{-2} = 2(2\ell+1)
\left( \begin{array}{cc} \ell & -1 \\ \ell(\ell-1) & -(\ell-1) \end{array} \right),
\end{equation}
and
\begin{equation}
 G_{0} = 2\left( \begin{array}{cc} \displaystyle \mu^2+\frac{2\omega^2-\sigma^2}{2\alpha_c^2}+ 4\ell a_2 +4\kappa_{\ell}a_0^2\delta_{\ell,1} & 
 \displaystyle -4a_2 +2\kappa_{\ell}a_0^2\delta_{\ell,1} \\ 
 \displaystyle 2\ell\mu^2+ 4\ell(\ell-1)a_2 +4\kappa_{\ell}a_0^2\delta_{\ell,1} & 
 \displaystyle -\mu^2+\frac{2\omega^2-\sigma^2}{2\alpha_c^2}- 4(\ell-1) a_2 +2\kappa_{\ell}a_0^2\delta_{\ell,1} \end{array} \right),\qquad
\end{equation}
it follows from the first relation in equation~(\ref{Eq:MatrixRels}) that
\begin{equation}
V_0 = b_1\left( \begin{array}{l} 1 \\ \ell \end{array} \right)
\end{equation}
for some real coefficient $b_1$, while the second relation in equation~(\ref{Eq:MatrixRels}) leads to
\begin{equation}
2(2\ell + 3)\left( \begin{array}{cc} -(\ell-1) & 1 \\ -\ell(\ell-1) & \ell \end{array} \right) V_2 
 = G_0 V_0.
\label{Eq:V2}
\end{equation}
The matrix on the left-hand side is not invertible, meaning that $G_0 V_0$ needs to lie in its image and that $V_2$ contains an additional free parameter. A careful calculation reveals that for all $\ell\geq 0$,
\begin{equation}
G_0 V_0 = b_1\left( 2\mu^2 + \frac{2\omega^2 - \sigma^2}{\alpha_c^2}
 + 12\kappa_\ell\delta_{\ell,1}a_0^2 \right)
 \left( \begin{array}{l} 1 \\ \ell \end{array} \right),
\end{equation}
which means that the system~(\ref{Eq:V2}) is solvable with
\begin{equation}
( 1-\ell, 1)\cdot V_2 = \frac{b_1}{2\ell + 3}\left( \mu^2 + \frac{2\omega^2 - \sigma^2}{2\alpha_c^2}
 + 6\kappa_\ell\delta_{\ell,1}a_0^2 \right).
\end{equation}
This finally leads to the following expansion near $r=0$ (setting $b_1 = 1$ without loss of generality)
\begin{subequations}\label{eq.right.boundary.coude}
\begin{eqnarray}
\delta\varphi_{\ell 1} &=& 1 + c r^2 + {\cal O}(r^4),
\label{Eq:DvarphiExp0}\\
\delta L &=& \ell + \left[ \frac{1}{2\ell + 3}\left( \mu^2 + \frac{2\omega^2 - \sigma^2}{2\alpha_c^2}
 + 6\kappa_\ell\delta_{\ell,1}a_0^2 \right) + (\ell-1) c \right] r^2 + {\cal O}(r^4),
\label{Eq:DLExp0}
\end{eqnarray}
\end{subequations}
with $c$ a free coefficient. For $\ell=0$ we may redefine
\begin{equation}
\frac{1}{3}\left( \mu^2 + \frac{2\omega^2 - \sigma^2}{2\alpha_c^2} \right) - c
  =: \frac{\mu^2}{\sigma(0)^2}\gamma,
\end{equation}
and the resulting expansion seems to agree with the one in equation~(38) in~\cite{Gleiser:1989a} (taking into account that $x = \mu r$ and $\sigma(0) = \sqrt{2\kappa_0}\psi_0(0)$).


\bibliographystyle{apsrev}
\bibliography{main}

\begin{thebibliography}{30}
\expandafter\ifx\csname natexlab\endcsname\relax\def\natexlab#1{#1}\fi
\expandafter\ifx\csname bibnamefont\endcsname\relax
  \def\bibnamefont#1{#1}\fi
\expandafter\ifx\csname bibfnamefont\endcsname\relax
  \def\bibfnamefont#1{#1}\fi
\expandafter\ifx\csname citenamefont\endcsname\relax
  \def\citenamefont#1{#1}\fi
\expandafter\ifx\csname url\endcsname\relax
  \def\url#1{\texttt{#1}}\fi
\expandafter\ifx\csname urlprefix\endcsname\relax\def\urlprefix{URL }\fi
\providecommand{\bibinfo}[2]{#2}
\providecommand{\eprint}[2][]{\url{#2}}

\bibitem[{\citenamefont{Kaup}(1968)}]{Kaup68}
\bibinfo{author}{\bibfnamefont{D.~J.} \bibnamefont{Kaup}},
  \bibinfo{journal}{Phys. Rev.} \textbf{\bibinfo{volume}{172}},
  \bibinfo{pages}{1331} (\bibinfo{year}{1968}).

\bibitem[{\citenamefont{Ruffini and Bonazzola}(1969)}]{Ruffini69}
\bibinfo{author}{\bibfnamefont{R.}~\bibnamefont{Ruffini}} \bibnamefont{and}
  \bibinfo{author}{\bibfnamefont{S.}~\bibnamefont{Bonazzola}},
  \bibinfo{journal}{Phys. Rev.} \textbf{\bibinfo{volume}{187}},
  \bibinfo{pages}{1767} (\bibinfo{year}{1969}).

\bibitem[{\citenamefont{Jetzer}(1992)}]{Jetzer92}
\bibinfo{author}{\bibfnamefont{P.}~\bibnamefont{Jetzer}},
  \bibinfo{journal}{Phys. Rep.} \textbf{\bibinfo{volume}{220}},
  \bibinfo{pages}{163} (\bibinfo{year}{1992}).

\bibitem[{\citenamefont{Schunck and Mielke}(2003)}]{Schunck:2003kk}
\bibinfo{author}{\bibfnamefont{F.~E.} \bibnamefont{Schunck}} \bibnamefont{and}
  \bibinfo{author}{\bibfnamefont{E.~W.} \bibnamefont{Mielke}},
  \bibinfo{journal}{Class. Quantum Grav.} \textbf{\bibinfo{volume}{20}},
  \bibinfo{pages}{R301} (\bibinfo{year}{2003}), \eprint{0801.0307}.

\bibitem[{\citenamefont{Liebling and Palenzuela}(2012)}]{Liebling:2012fv}
\bibinfo{author}{\bibfnamefont{S.~L.} \bibnamefont{Liebling}} \bibnamefont{and}
  \bibinfo{author}{\bibfnamefont{C.}~\bibnamefont{Palenzuela}},
  \bibinfo{journal}{Living Rev.Rel.} \textbf{\bibinfo{volume}{15}},
  \bibinfo{pages}{6} (\bibinfo{year}{2012}), \eprint{1202.5809}.

\bibitem[{\citenamefont{{Schunck} and {Mielke}}(1996)}]{1996rscc.conf..138S}
\bibinfo{author}{\bibfnamefont{F.~E.} \bibnamefont{{Schunck}}}
  \bibnamefont{and} \bibinfo{author}{\bibfnamefont{E.~W.}
  \bibnamefont{{Mielke}}}, in \emph{\bibinfo{booktitle}{Relativity and
  Scientific Computing. Computer Algebra, Numerics, Visualization}}
  (\bibinfo{year}{1996}), pp. \bibinfo{pages}{138--151}.

\bibitem[{\citenamefont{Yoshida and Eriguchi}(1997)}]{Yoshida:1997qf}
\bibinfo{author}{\bibfnamefont{S.}~\bibnamefont{Yoshida}} \bibnamefont{and}
  \bibinfo{author}{\bibfnamefont{Y.}~\bibnamefont{Eriguchi}},
  \bibinfo{journal}{Phys. Rev. D} \textbf{\bibinfo{volume}{56}},
  \bibinfo{pages}{762} (\bibinfo{year}{1997}).

\bibitem[{\citenamefont{Palenzuela et~al.}(2017)\citenamefont{Palenzuela, Pani,
  Bezares, Cardoso, Lehner, and Liebling}}]{Palenzuela:2017kcg}
\bibinfo{author}{\bibfnamefont{C.}~\bibnamefont{Palenzuela}},
  \bibinfo{author}{\bibfnamefont{P.}~\bibnamefont{Pani}},
  \bibinfo{author}{\bibfnamefont{M.}~\bibnamefont{Bezares}},
  \bibinfo{author}{\bibfnamefont{V.}~\bibnamefont{Cardoso}},
  \bibinfo{author}{\bibfnamefont{L.}~\bibnamefont{Lehner}}, \bibnamefont{and}
  \bibinfo{author}{\bibfnamefont{S.}~\bibnamefont{Liebling}},
  \bibinfo{journal}{Phys. Rev. D} \textbf{\bibinfo{volume}{96}},
  \bibinfo{pages}{104058} (\bibinfo{year}{2017}), \eprint{1710.09432}.

\bibitem[{\citenamefont{Seidel and Suen}(1994)}]{Seidel:1993zk}
\bibinfo{author}{\bibfnamefont{E.}~\bibnamefont{Seidel}} \bibnamefont{and}
  \bibinfo{author}{\bibfnamefont{W.-M.} \bibnamefont{Suen}},
  \bibinfo{journal}{Phys. Rev. Lett.} \textbf{\bibinfo{volume}{72}},
  \bibinfo{pages}{2516} (\bibinfo{year}{1994}), \eprint{gr-qc/9309015}.

\bibitem[{\citenamefont{Sanchis-Gual et~al.}(2019)\citenamefont{Sanchis-Gual,
  Di~Giovanni, Zilh\~ao, Herdeiro, Cerd\'a-Dur\'an, Font, and
  Radu}}]{Sanchis-Gual:2019ljs}
\bibinfo{author}{\bibfnamefont{N.}~\bibnamefont{Sanchis-Gual}},
  \bibinfo{author}{\bibfnamefont{F.}~\bibnamefont{Di~Giovanni}},
  \bibinfo{author}{\bibfnamefont{M.}~\bibnamefont{Zilh\~ao}},
  \bibinfo{author}{\bibfnamefont{C.}~\bibnamefont{Herdeiro}},
  \bibinfo{author}{\bibfnamefont{P.}~\bibnamefont{Cerd\'a-Dur\'an}},
  \bibinfo{author}{\bibfnamefont{J.~A.} \bibnamefont{Font}}, \bibnamefont{and}
  \bibinfo{author}{\bibfnamefont{E.}~\bibnamefont{Radu}},
  \bibinfo{journal}{Phys. Rev. Lett.} \textbf{\bibinfo{volume}{123}},
  \bibinfo{pages}{221101} (\bibinfo{year}{2019}), \eprint{1907.12565}.

\bibitem[{\citenamefont{Alcubierre et~al.}(2018)\citenamefont{Alcubierre,
  Barranco, Bernal, Degollado, Diez-Tejedor, Megevand, Nunez, and
  Sarbach}}]{Alcubierre:2018ahf}
\bibinfo{author}{\bibfnamefont{M.}~\bibnamefont{Alcubierre}},
  \bibinfo{author}{\bibfnamefont{J.}~\bibnamefont{Barranco}},
  \bibinfo{author}{\bibfnamefont{A.}~\bibnamefont{Bernal}},
  \bibinfo{author}{\bibfnamefont{J.~C.} \bibnamefont{Degollado}},
  \bibinfo{author}{\bibfnamefont{A.}~\bibnamefont{Diez-Tejedor}},
  \bibinfo{author}{\bibfnamefont{M.}~\bibnamefont{Megevand}},
  \bibinfo{author}{\bibfnamefont{D.}~\bibnamefont{Nunez}}, \bibnamefont{and}
  \bibinfo{author}{\bibfnamefont{O.}~\bibnamefont{Sarbach}},
  \bibinfo{journal}{Class. Quant. Grav.} \textbf{\bibinfo{volume}{35}},
  \bibinfo{pages}{19LT01} (\bibinfo{year}{2018}), \eprint{1805.11488}.

\bibitem[{\citenamefont{Olabarrieta et~al.}(2007)\citenamefont{Olabarrieta,
  Ventrella, Choptuik, and Unruh}}]{Olabarrieta:2007di}
\bibinfo{author}{\bibfnamefont{I.}~\bibnamefont{Olabarrieta}},
  \bibinfo{author}{\bibfnamefont{J.~F.} \bibnamefont{Ventrella}},
  \bibinfo{author}{\bibfnamefont{M.~W.} \bibnamefont{Choptuik}},
  \bibnamefont{and} \bibinfo{author}{\bibfnamefont{W.~G.} \bibnamefont{Unruh}},
  \bibinfo{journal}{Phys. Rev.} \textbf{\bibinfo{volume}{D76}},
  \bibinfo{pages}{124014} (\bibinfo{year}{2007}), \eprint{0708.0513}.

\bibitem[{\citenamefont{Gleiser}(1988)}]{Gleiser:1988rq}
\bibinfo{author}{\bibfnamefont{M.}~\bibnamefont{Gleiser}},
  \bibinfo{journal}{Phys. Rev.} \textbf{\bibinfo{volume}{D38}},
  \bibinfo{pages}{2376} (\bibinfo{year}{1988}), \bibinfo{note}{[Erratum: Phys.
  Rev.D39,no.4,1257(1989)]}.

\bibitem[{\citenamefont{Gleiser and Watkins}(1989)}]{Gleiser:1989a}
\bibinfo{author}{\bibfnamefont{M.}~\bibnamefont{Gleiser}} \bibnamefont{and}
  \bibinfo{author}{\bibfnamefont{R.}~\bibnamefont{Watkins}},
  \bibinfo{journal}{Nucl. Phys.} \textbf{\bibinfo{volume}{B319}},
  \bibinfo{pages}{733} (\bibinfo{year}{1989}), \eprint{gr-qc/9905067}.

\bibitem[{\citenamefont{T.~D.~Lee}(1989)}]{Lee89}
\bibinfo{author}{\bibfnamefont{P.~Y.} \bibnamefont{T.~D.~Lee}},
  \bibinfo{journal}{Nucl. Phys} \textbf{\bibinfo{volume}{B315}},
  \bibinfo{pages}{447} (\bibinfo{year}{1989}).

\bibitem[{\citenamefont{Balakrishna et~al.}(1998)\citenamefont{Balakrishna,
  Seidel, and Suen}}]{Balakrishna:1997ej}
\bibinfo{author}{\bibfnamefont{J.}~\bibnamefont{Balakrishna}},
  \bibinfo{author}{\bibfnamefont{E.}~\bibnamefont{Seidel}}, \bibnamefont{and}
  \bibinfo{author}{\bibfnamefont{W.-M.} \bibnamefont{Suen}},
  \bibinfo{journal}{Phys. Rev. D} \textbf{\bibinfo{volume}{58}},
  \bibinfo{pages}{104004} (\bibinfo{year}{1998}), \eprint{gr-qc/9712064}.

\bibitem[{\citenamefont{Seidel and Suen}(1990)}]{Seidel90}
\bibinfo{author}{\bibfnamefont{E.}~\bibnamefont{Seidel}} \bibnamefont{and}
  \bibinfo{author}{\bibfnamefont{W.}~\bibnamefont{Suen}},
  \bibinfo{journal}{Phys. Rev.} \textbf{\bibinfo{volume}{D42}},
  \bibinfo{pages}{384} (\bibinfo{year}{1990}).

\bibitem[{\citenamefont{Hawley and Choptuik}(2000)}]{Hawley2000}
\bibinfo{author}{\bibfnamefont{S.}~\bibnamefont{Hawley}} \bibnamefont{and}
  \bibinfo{author}{\bibfnamefont{M.}~\bibnamefont{Choptuik}},
  \bibinfo{journal}{Phys. Rev.} \textbf{\bibinfo{volume}{D62}},
  \bibinfo{pages}{104024} (\bibinfo{year}{2000}), \eprint{gr-qc/0007039}.

\bibitem[{\citenamefont{Guzman}(2009)}]{Guzman09}
\bibinfo{author}{\bibfnamefont{F.}~\bibnamefont{Guzman}},
  \bibinfo{journal}{Revista Mexicana de Fisica} \textbf{\bibinfo{volume}{55}},
  \bibinfo{pages}{321} (\bibinfo{year}{2009}).

\bibitem[{\citenamefont{Kusmartsev et~al.}(1991)\citenamefont{Kusmartsev,
  Mielke, and Schunck}}]{Kusmartsev:1990cr}
\bibinfo{author}{\bibfnamefont{F.~V.} \bibnamefont{Kusmartsev}},
  \bibinfo{author}{\bibfnamefont{E.~W.} \bibnamefont{Mielke}},
  \bibnamefont{and} \bibinfo{author}{\bibfnamefont{F.~E.}
  \bibnamefont{Schunck}}, \bibinfo{journal}{Phys. Rev. D}
  \textbf{\bibinfo{volume}{43}}, \bibinfo{pages}{3895} (\bibinfo{year}{1991}),
  \eprint{0810.0696}.

\bibitem[{\citenamefont{Chandrasekhar}(1964{\natexlab{a}})}]{Chandrasekhar:1964zza}
\bibinfo{author}{\bibfnamefont{S.}~\bibnamefont{Chandrasekhar}},
  \bibinfo{journal}{Phys. Rev. Lett.} \textbf{\bibinfo{volume}{12}},
  \bibinfo{pages}{114} (\bibinfo{year}{1964}{\natexlab{a}}).

\bibitem[{\citenamefont{Chandrasekhar}(1964{\natexlab{b}})}]{Chandrasekhar:1964zz}
\bibinfo{author}{\bibfnamefont{S.}~\bibnamefont{Chandrasekhar}},
  \bibinfo{journal}{Astrophys. J.} \textbf{\bibinfo{volume}{140}},
  \bibinfo{pages}{417} (\bibinfo{year}{1964}{\natexlab{b}}),
  \bibinfo{note}{[Erratum: Astrophys.J. 140, 1342 (1964)]}.

\bibitem[{\citenamefont{Shapiro and Teukolsky}(1983)}]{Shapiro:1983du}
\bibinfo{author}{\bibfnamefont{S.~L.} \bibnamefont{Shapiro}} \bibnamefont{and}
  \bibinfo{author}{\bibfnamefont{S.~A.} \bibnamefont{Teukolsky}},
  \emph{\bibinfo{title}{{Black holes, white dwarfs, and neutron stars: The
  physics of compact objects}}} (\bibinfo{year}{1983}), ISBN
  \bibinfo{isbn}{978-0-471-87316-7}.

\bibitem[{\citenamefont{Alcubierre et~al.}(2019)\citenamefont{Alcubierre,
  Barranco, Bernal, Degollado, Diez-Tejedor, Megevand, N\'u\~nez, and
  Sarbach}}]{Alcubierre:2019qnh}
\bibinfo{author}{\bibfnamefont{M.}~\bibnamefont{Alcubierre}},
  \bibinfo{author}{\bibfnamefont{J.}~\bibnamefont{Barranco}},
  \bibinfo{author}{\bibfnamefont{A.}~\bibnamefont{Bernal}},
  \bibinfo{author}{\bibfnamefont{J.~C.} \bibnamefont{Degollado}},
  \bibinfo{author}{\bibfnamefont{A.}~\bibnamefont{Diez-Tejedor}},
  \bibinfo{author}{\bibfnamefont{M.}~\bibnamefont{Megevand}},
  \bibinfo{author}{\bibfnamefont{D.}~\bibnamefont{N\'u\~nez}},
  \bibnamefont{and} \bibinfo{author}{\bibfnamefont{O.}~\bibnamefont{Sarbach}},
  \bibinfo{journal}{Class. Quant. Grav.} \textbf{\bibinfo{volume}{36}},
  \bibinfo{pages}{215013} (\bibinfo{year}{2019}), \eprint{1906.08959}.

\bibitem[{\citenamefont{Jaramillo et~al.}(2020)\citenamefont{Jaramillo,
  Sanchis-Gual, Barranco, Bernal, Degollado, Herdeiro, and
  N\'u\~nez}}]{Jaramillo:2020rsv}
\bibinfo{author}{\bibfnamefont{V.}~\bibnamefont{Jaramillo}},
  \bibinfo{author}{\bibfnamefont{N.}~\bibnamefont{Sanchis-Gual}},
  \bibinfo{author}{\bibfnamefont{J.}~\bibnamefont{Barranco}},
  \bibinfo{author}{\bibfnamefont{A.}~\bibnamefont{Bernal}},
  \bibinfo{author}{\bibfnamefont{J.~C.} \bibnamefont{Degollado}},
  \bibinfo{author}{\bibfnamefont{C.}~\bibnamefont{Herdeiro}}, \bibnamefont{and}
  \bibinfo{author}{\bibfnamefont{D.}~\bibnamefont{N\'u\~nez}},
  \bibinfo{journal}{Phys. Rev. D} \textbf{\bibinfo{volume}{101}},
  \bibinfo{pages}{124020} (\bibinfo{year}{2020}), \eprint{2004.08459}.

\bibitem[{\citenamefont{Sanchis-Gual et~al.}(2021)\citenamefont{Sanchis-Gual,
  Di~Giovanni, Herdeiro, Radu, and Font}}]{Sanchis-Gual:2021edp}
\bibinfo{author}{\bibfnamefont{N.}~\bibnamefont{Sanchis-Gual}},
  \bibinfo{author}{\bibfnamefont{F.}~\bibnamefont{Di~Giovanni}},
  \bibinfo{author}{\bibfnamefont{C.}~\bibnamefont{Herdeiro}},
  \bibinfo{author}{\bibfnamefont{E.}~\bibnamefont{Radu}}, \bibnamefont{and}
  \bibinfo{author}{\bibfnamefont{J.~A.} \bibnamefont{Font}}
  (\bibinfo{year}{2021}), \eprint{gr-qc/2103.12136}.

\bibitem[{\citenamefont{Amann and Quittner}(1995)}]{hApQ95}
\bibinfo{author}{\bibfnamefont{H.}~\bibnamefont{Amann}} \bibnamefont{and}
  \bibinfo{author}{\bibfnamefont{P.}~\bibnamefont{Quittner}},
  \bibinfo{journal}{J. Math. Phys.} \textbf{\bibinfo{volume}{36}},
  \bibinfo{pages}{4553} (\bibinfo{year}{1995}).

\bibitem[{\citenamefont{Reed and Simon}(1980{\natexlab{a}})}]{ReedSimon80II}
\bibinfo{author}{\bibfnamefont{M.}~\bibnamefont{Reed}} \bibnamefont{and}
  \bibinfo{author}{\bibfnamefont{B.}~\bibnamefont{Simon}},
  \emph{\bibinfo{title}{Methods of Modern Mathematical Physics, Vol. II:
  Fourier Analysis, Self-Adjointness}} (\bibinfo{publisher}{Academic Press},
  \bibinfo{address}{San Diego}, \bibinfo{year}{1980}{\natexlab{a}}).

\bibitem[{\citenamefont{Reed and Simon}(1980{\natexlab{b}})}]{ReedSimon80I}
\bibinfo{author}{\bibfnamefont{M.}~\bibnamefont{Reed}} \bibnamefont{and}
  \bibinfo{author}{\bibfnamefont{B.}~\bibnamefont{Simon}},
  \emph{\bibinfo{title}{Methods of Modern Mathematical Physics, Vol. I:
  Functional Analysis}} (\bibinfo{publisher}{Academic Press},
  \bibinfo{address}{San Diego}, \bibinfo{year}{1980}{\natexlab{b}}).

\bibitem[{\citenamefont{Reed and Simon}(1980{\natexlab{c}})}]{ReedSimon80IV}
\bibinfo{author}{\bibfnamefont{M.}~\bibnamefont{Reed}} \bibnamefont{and}
  \bibinfo{author}{\bibfnamefont{B.}~\bibnamefont{Simon}},
  \emph{\bibinfo{title}{Methods of Modern Mathematical Physics, Vol. IV:
  Analysis of Operators}} (\bibinfo{publisher}{Academic Press},
  \bibinfo{address}{San Diego}, \bibinfo{year}{1980}{\natexlab{c}}).

\end{thebibliography}


\end{document}